\documentstyle[prd,aps,preprint,tighten,epsf]{revtex}

\def\Psibar{\overline{\Psi}}
\def\psibar{\overline{\psi}}
\def\w3m{\left\langle\left|W\right|\right\rangle}
\def\pbp{\left\langle\overline{\psi}\psi\right\rangle}
\def\plaq{\langle{\rm plaq}\rangle}
\begin{document}
%
\preprint{CU-TP-938}
\title{The finite temperature QCD phase transition with domain wall fermions}
\author{P.~Chen,
N.~Christ,
G.~Fleming,
A.~Kaehler,
C.~Malureanu,
R.~Mawhinney,
G.~Siegert,
C.~Sui,
L.~Wu,
Y.~Zhestkov}
\address{Physics Department,
Columbia University,
New York, NY 10027}
\author{P.~Vranas}
\address{Physics Department,
University of Illinois,
Urbana, IL 61801}

\date{June 6, 2000}
\maketitle

\begin{abstract}
The domain wall formulation of lattice fermions is expected to
support accurate chiral symmetry, even at finite lattice spacing.  Here
we attempt to use this new fermion formulation to simulate two-flavor,
finite temperature QCD near the chiral phase transition.  In this initial
study, a variety of quark masses, domain wall heights and domain wall 
separations are explored using an $8^3 \times 4$ lattice.  Both the 
expectation value of the Wilson line and the chiral condensate show
the temperature dependence expected for the QCD phase transition.
Further, the desired chiral properties are seen for the chiral condensate, 
suggesting that the domain wall fermion formulation may be an effective 
approach for the numerical study of QCD at finite temperature.
\end{abstract}


\section{Introduction}
\label{sec:intro}

Many of the properties of low energy QCD are a direct consequence
of the breaking of chiral symmetry by the QCD vacuum.  It is expected
that this spontaneous chiral symmetry breaking will disappear as the
temperature is increased.  Both the nature of this symmetry restoration
(abrupt phase transition or continuous cross-over) and the character
of the high-temperature quark-gluon plasma phase are active areas of 
both theoretical\cite{Bass:1999zq,Karsch:1999vy} and experimental 
research\cite{Kluberg:1999xx,Stock:1999mr}.

An especially promising approach to the theoretical study of 
equilibrium properties of both the QCD phase transition and the 
high-temperature plasma phase is direct numerical simulation 
of the Feynman path integral using the methods of lattice 
gauge theory.  The quantum partition function is written as a 
Euclidean path integral that can be studied {\it ab initio} 
using the discrete, lattice formulation of Wilson\cite{Wilson:1975id}.
While the local color gauge symmetry of the theory remains exact 
at any lattice spacing in Wilson's formulation, much of the
theory's flavor symmetry, and especially its chiral component, 
is explicitly broken. 

This difficulty in representing the continuum flavor symmetries in
a lattice fermion formulation is a serious problem that has
persisted for more than two decades.  When the fermion action is 
naively discretized the low-energy fermionic degrees of freedom 
increase by a factor of $2^4$.  This well-known ``doubling'' problem 
can only be remedied by methods that explicitly break the chiral 
flavor symmetries for finite lattice spacing \cite{Nielsen:1981hk}. 
The chiral symmetries are then recovered together with the Lorentz 
symmetry as the lattice spacing is sent to zero. The most popular 
of these methods are 
staggered\cite{Kogut:1975ag,Banks:1976gq,Susskind:1977jm} and 
Wilson\cite{Wilson:1975id} fermions.  

Although, in principle these methods should be able to approximate 
the continuum theory in a controlled way, in practice this problem 
has been a formidable obstacle to lattice studies of the QCD phase 
transition.  For example, the Wilson fermion formulation explicitly 
breaks all of the continuum chiral symmetries making phenomena 
driven by the spontaneous breakdown of chiral symmetry difficult to
study.  While staggered fermions do possess a one-dimensional 
continuous chiral symmetry at finite lattice spacing, this formulation
explicitly breaks the vector flavor symmetry so instead of three 
light Goldstone pions with mass on the order of the critical 
temperature $T_c \approx 160$ MeV as found in Nature, present 
staggered simulations have masses for two of the three pions in the 
range 500-600 GeV, certainly too large.  

In addition, the subtle effects of the continuum axial anomaly which 
are closely connected with the order of the 
transition\cite{Pisarski:1984ms} are badly mutilated by both 
fermion formalisms at finite lattice spacing.  While the anomalous 
$U_A(1)$ continuum chiral symmetry is explicitly broken by both 
formalisms, the fermion zero modes required by Atiyah-Singer
index theorem are shifted away from zero by finite lattice spacing 
effects.

In principle, each of these difficulties can be addressed by simply 
working at smaller lattice spacing.  However, present numerical 
methods scale very poorly as the lattice spacing is decreased, 
with the required numerical effort growing as $\sim 1/a^{8-10}$
for lattice spacing $a$.

Domain wall fermions (DWF) offer a new approach to the problem of
including fermions in lattice gauge theory calculations.  In this 
formulation, introduced by Kaplan \cite{Kaplan:1992bt,Kaplan:1993sg}, 
the fermionic fields are defined on a five-dimensional hyper-cubic 
lattice using a local action.  The fifth direction can be thought 
of as an extra space-time dimension or as a new internal flavor 
space.  The gauge fields are represented in the standard way in 
four dimensional space-time and are coupled to the extra fermion 
degrees of freedom in a diagonal fashion.

In this paper, we use a variant of Kaplan's approach, developed by 
Shamir\cite{Shamir:1993zy}, in which the two four-dimensional 
faces orthogonal to the new fifth dimension are treated differently, 
with free boundary conditions imposed on the fermion fields.  This 
key ingredient allows a system made up of naively massive fermions 
to develop chiral surface states on these boundaries (domain walls) 
with the positive chirality states bound exponentially to one wall 
and the negative chirality states bound to the other.  

The two chiralities overlap only by an amount that is exponentially 
small in $L_s$, the number of lattice sites along 
the fifth direction.  The resulting mixed state forms a Dirac 
4-spinor that propagates in the four-dimensional space-time with 
an exponentially small mass.  Therefore, the amount of chiral 
symmetry breaking that is artificially induced by the regulator 
can be controlled by the new parameter $L_s$.  In the 
$L_s \rightarrow \infty$ limit the chiral symmetry is exact even 
at finite lattice spacing.  Thus, the domain wall fermion method has 
succeeded in disentangling the chiral limit ($L_s\rightarrow \infty$) 
and the continuum limit ($a \rightarrow 0$).  Furthermore, the direct 
computing requirement grows only linearly with $L_s$.

Here we report the first full QCD simulations using domain wall 
fermions in four dimensions.  The properties and parameter space
of domain wall fermions appropriate for a study of QCD 
thermodynamics are explored in detail.  Small lattices of size $8^3
\times 4$ were used to perform numerical simulations of full, 
two-flavor QCD at finite temperature.   Preliminary results of this
work have appeared in \cite{Vranas:1998vm,Chen:1998xw,Vranas:1999pm}.
These studies have been carried out using the QCDSP supercomputer at 
Columbia\cite{Chen:1998cg}.  Based on the work reported here,
results of physical interest have been obtained on larger 
lattices for a variety of observables.  Preliminary results of these
studies can be found in \cite{Vranas:1998vm,Vranas:1999pm} and will be 
presented in follow-on papers\cite{Columbia_cur}.

For a detailed introduction to the subject and relevant references 
the reader is referred to 
Refs.~\cite{Vranas:1997tj,Vranas:1997da,Chen:1998ne}, and the reviews in
Refs.~\cite{Narayanan:1993gq,Creutz:1994px,Shamir:1995zx,Blum:1998ud}.  
Earlier numerical work using domain wall fermions has explored the parameter 
space of a QCD-like, dynamical vector theory in two dimensions, the two flavor 
Schwinger model\cite{Vranas:1997tj,Vranas:1997da}.  For applications 
to quenched QCD see
Refs.~\cite{Blum:1997mz,Blum:1997jf,Mawhinney:1998ut,Fleming:1998cc,Kaehler:1998xx,Edwards:1998sh,Fleming:1999eq,Wu:1999cd,AliKhan:1999zn,Edwards:1999bm,Aoki:1999uv,Aoki:2000pc,Chen:1998xw} for applications to four-Fermi models see
Ref.~\cite{Vranas:1999nx} and for possible alternatives to domain wall 
fermion simulations see 
Refs.~\cite{Neuberger:1998bg,Neuberger:1997fp,Neuberger:1998my,Edwards:1998wx,Liu:1998hj,Edwards:2000qv,Narayanan:2000qx}.

In Section~\ref{sec:hmc} the action of the theory and a brief
description of the numerical methods are presented. In 
Section~\ref{sec:analytical} some important analytical facts are 
outlined in order to help guide the numerical investigation.  
In Section~\ref{sec:trans} we study the chiral properties of the
theory both below and above the chiral phase transition.  Our 
numerical results suggest that domain wall fermions are able to 
sustain the desired chiral properties of QCD, even at finite 
lattice spacing.  Both a low temperature phase where the 
$SU(2) \times SU(2)$ chiral symmetry is broken spontaneously 
to an $SU(2)$ vector symmetry and a high temperature phase where 
the full $SU(2) \times SU(2)$ chiral symmetry is intact can be 
recognized.

In Section \ref{sec:trans} the dependence on the two new regulator 
parameters, the number of sites in the fifth direction $L_s$, 
and the domain wall ``height'' $m_0$, is studied numerically.  Finally,
in Section~\ref{sec:conclusion} conclusions and outlook are presented.
Appendix~\ref{appendix_gamma} gives the explicit form of the 
gamma matrices used in this work while Appendix~B describes the 
molecular dynamics equations of motion.  Tables summarizing the 
numerical results are given at the end of the paper.

\section{Hybrid Monte Carlo with domain wall fermions}
\label{sec:hmc}

In this section the action of QCD with domain wall fermions,
its implementation for the Hybrid Monte Carlo (HMC) algorithm,
and the parameters used in the simulations are described.
In the following, we discuss the case of two degenerate flavors 
implemented using the HMC $\Phi$ algorithm\cite{Gottlieb:1987mq}.  
(An odd number of flavors can be simulated using the HMC $R$ 
algorithm\cite{Gottlieb:1987mq}).

Domain wall fermions can be used in numerical simulations in a fashion
similar to traditional Wilson fermions. In fact, if the fifth direction
is thought of as an internal flavor direction then an HMC simulation
with DWF is identical to a simulation of many flavors of Wilson
fermions with a sophisticated mass matrix.  We use the partition 
function of QCD with domain wall fermions proposed in 
\cite{Furman:1995ky} but with a slightly different heavy flavor 
subtraction as in \cite{Vranas:1997tj,Vranas:1997da}.  In particular:
\begin{equation}
Z = \int [dU] \int [d\Psibar d\Psi] 
        \int [d\Phi_{PV}^\dagger d\Phi_{PV}] e^{-S}
\label{Z}
\end{equation}
$U_\mu(x)$ is the gauge field, $\Psi(x,s)$ is the fermion field and
$\Phi_{PV}(x,s)$ is a bosonic, Pauli-Villars field.  The
variable $x$ specifies the coordinates in the four-dimensional 
space-time box with extent $L$ along each of the spatial directions 
and extent $N_t$ along the time direction while $s=0,1, \dots, L_s-1$ 
is the coordinate of the fifth direction, with $L_s$ assumed to be even.  
The action $S$ is given by:
\begin{equation}
S = S_G(U) + S_F(\Psibar, \Psi, U) +
S_{PV}(\Phi_{PV}^\dagger, \Phi_{PV}, U)
\label{action}
\end{equation}
where:
\begin{equation}
S_G = \beta \sum_p ( 1 - {1 \over 3} {\rm Re Tr}[U_p])
\label{action_G}
\end{equation}
is the standard plaquette action, $\beta = 6/ g_0^2$ and $g_0$ is the
lattice gauge coupling. The fermion action for two flavors is:
\begin{equation}
S_F = - \sum_{x,x^\prime,s,s^\prime,f} \Psibar_f(x,s) D_F(x,s; x^\prime,
s^\prime) \Psi_f(x^\prime,s^\prime)
\label{action_F}
\end{equation}
with flavor index $f=1,2$ and Dirac operator:
\begin{equation}
D_F(x,s; x^\prime, s^\prime) = \delta_{s,s^\prime} D^\parallel(x,x^\prime)
+ D^\bot(s,s^\prime) \delta_{x,x^\prime}
\label{eq:D_F}
\end{equation}
\begin{eqnarray}
D^\parallel(x,x^\prime) &=& {1\over 2} \sum_{\mu=1}^4 \left[ (1-\gamma_\mu)
U_\mu(x) \delta_{x+\hat\mu,x^\prime} + (1+\gamma_\mu)
U^\dagger_\mu(x^\prime) \delta_{x-\hat\mu,x^\prime} \right] \nonumber \\
&+& (m_0 - 4)\delta_{x,x^\prime}
\label{eq:D_parallel}
\end{eqnarray}
\begin{eqnarray}
D^\bot(s,s^\prime) 
     &=& {1\over 2}\Big[(1-\gamma_5)\delta_{s+1,s^\prime} 
                 + (1+\gamma_5)\delta_{s-1,s^\prime} 
                 - 2\delta_{s,s^\prime}\Big] \nonumber\\
     &-& {m_f\over 2}\Big[(1-\gamma_5)\delta_{L_s-1,s}\delta_{s^\prime,0}
      +  (1+\gamma_5)\delta_{s,0}\delta_{L_s-1,s^\prime}\Big]
\end{eqnarray}
\label{D_perp}
Here, $s$ and $s^\prime$ lie in the range $0 \le s,s^\prime \le L_s-1$.
In the above equations $m_0$ is a five-dimensional mass representing
the height of the domain wall in Kaplan's original language. 
In order for the doubler species to be removed in the free theory one 
must choose $0<m_0<2$\cite{Kaplan:1992bt,Kaplan:1993sg}.  The parameter 
$m_f$ explicitly mixes the two chiralities and, as a result, controls 
the bare fermion mass of the four-dimensional effective theory.

While the DWF Dirac operator defined above is not hermitian, it does
obey the identity \cite{Furman:1995ky}:
\begin{equation}
\gamma_5 R_5 D_F \gamma_5 R_5 = D_F^\dagger
\label{reflection}
\end{equation}
with $R_5$ the reflection operator along the fifth direction. 
As a result the single-flavor Dirac determinant is real:
$\det{D_F}^*=\det{D_F^\dagger}=\det\gamma_5 R_5 D_F \gamma_5 R_5=\det{D_F}$
and the two-flavor determinant which follows from integrating
out the fermions in Eq.~\ref{Z}, $\det{D_F}^2$, is positive.
The gamma matrices used in this work are given in 
Appendix~\ref{appendix_gamma}.   Also notice that $D_F$ is the same
as the $D_F^\dagger$ of \cite{Furman:1995ky}.

The Pauli-Villars action is designed to cancel the contribution of 
the heavy fermions in the large $L_s$ limit.  Normally, such heavy 
fermions decouple from low energy physics and can be safely ignored.  
However, in the present situation the number of heavy fermions grows 
proportional to $L_s$ and can potentially overwhelm the effects of 
the fixed number of low energy degrees of freedom of interest.   
Specifically this difficulty will arise for the order of limits for 
which DWF are intended: first $L_s\rightarrow\infty$ followed by 
$a \rightarrow 0$.
\cite{Narayanan:1993wx,Narayanan:1993ss,Narayanan:1994sk,Narayanan:1995gw}.

There is some flexibility in the definition of the Pauli-Villars 
action since different actions can easily have the same 
$L_s \rightarrow \infty$ limit.  However, the choice of the 
Pauli-Villars action may affect the approach to the $L_s
\rightarrow \infty$ limit.   A slightly different action than 
that proposed by Furman and Shamir\cite{Furman:1995ky} is used 
here.  This action \cite{Vranas:1997tj,Vranas:1997da} is easier to 
implement numerically and, even for finite $L_s$, it exactly cancels 
the fermion action when $m_f = 1$ resulting in a pure gauge theory.  
For two fermion flavors, the Pauli-Villars action we use is:
\begin{equation}
S_{PV} =
\sum_{x,x^\prime,s,s^\prime} \Phi_{PV}^\dagger(x,s) 
M_F(x,s; x^\prime, s^\prime)|_{m_f=1} \Phi_{PV}(x^\prime,s^\prime)
\label{action_PV}
\end{equation}
where $M_F=D_F^\dagger D_F$.

The traditional HMC $\Phi$ algorithm was constructed directly from 
the action of Eq.~\ref{action}.   In order to improve performance a 
standard even-odd preconditioning\cite{DeGrand:1988vx} of the Dirac 
operator $D_F$ was employed. The even-odd preconditioning was done 
on the five dimensional space.  All necessary matrix inversions were 
done using a standard conjugate gradient (CG) algorithm. As expected 
the even-odd preconditioning resulted in a reduction of the 
required number of conjugate gradient iterations and a consequent 
speed-up of a factor of approximately two.  

The only new ingredient in our HMC algorithm is the appearance 
of the bosonic Pauli-Villars fields.  The probability distribution 
of these fields is generated with a heat bath step at the 
beginning of each HMC ``trajectory'': a field of Gaussian 
random numbers is generated with distribution $e^{-\eta_{PV}^\dagger \eta_{PV}}$ 
and from it the Pauli-Villars fields $\Phi_{PV}(x,s)$ are obtained by 
$\Phi_{PV}=[D_F(m_f=1)]^{-1} \eta_{PV}$ using the CG algorithm.  

Since the Pauli-Villars action in Eq.~\ref{action_PV} is 
polynomial in the domain wall operator $D_F$, its gradient with 
respect to the gauge fields, needed to evolve the gauge degrees 
of freedom, can be computed without performing any Dirac 
inversions.  This contrasts favorably with the fermion 
contribution to the gauge force which requires one inversion 
per molecular dynamics step. As a result, the relative 
computational cost involved in calculating the Pauli-Villars 
force is negligible.  Furthermore, because the Pauli-Villars 
fields are bosonic their molecular dynamics force term enters 
with an opposite sign that of the fermion force, resulting 
in a large, approximate cancellation.  Because of this cancelation 
the HMC force term is approximately independent of $L_s$ and it is 
not necessary to decrease the HMC step size as $L_s$ is increased.

In the approach described above the presence of the Pauli-Villars 
fields increases the memory requirement.  However, it
should be noted that there is an alternative approach that does 
not involve Pauli-Villars fields.  To see this consider the result 
after integration over both the Pauli-Villars and fermion fields. 
It is $\det{M_F(m_f)} / \det{M_F(m_f=1)} =
\det{[M_F(m_f) / M_F(m_f=1)]}$.   Therefore, one could simulate the 
same action without Pauli-Villars fields by simply using as the 
fermion matrix $M_F(m_f) / M_F(m_f=1)$.  Inversion of this matrix 
will involve inversion of $M_F(m_f)$ using the CG algorithm as 
in the previous method while the final result would have to be 
multiplied by the matrix $M_F(m_f=1)$.  If, for example, the CG 
algorithm required $100$ iterations to converge, this extra matrix 
multiplication will increase the computing cost by only $1\%$.  
The only disadvantage of this approach is that the equations of 
motion become slightly more complicated. 

Since this work is the first to implement DWF in dynamical QCD 
the approach with Pauli-Villars fields was adopted for 
simplicity and because it has been proven reliable in 
numerical simulations of the Schwinger 
model\cite{Vranas:1997tj,Vranas:1997da}. For the convenience of the 
reader the molecular dynamics equations of motion with 
Pauli-Villars fields and an even-odd preconditioned 
DWF Dirac operator are given in Appendix~\ref{appendix_md}.

Fermionic Green's functions were computed using the method described in
Ref.~\cite{Furman:1995ky}.  Standard fermion fields in the four-dimensional
space--time are constructed from the five-dimensional fermion fields
using the projection prescription:
\begin{eqnarray} 
\psi(x)    &=& P_{\rm L} \Psi(x,0) + P_{\rm R} \Psi(x, L_s-1) \nonumber \\
\psibar(x) &=& \Psibar(x,L_s-1) P_{\rm L} + \Psibar(x, 0) P_{\rm R}
\label{projection}
\end{eqnarray} 
where $P_{\rm R/L}={1\over 2}(1\pm\gamma^5)$.  This somewhat arbitrary
choice defines operators which should have a large overlap with the 
physical low energy fermion modes bound to the $s=0$ and $s=L_s-1$ walls.
The right- and left-handed components found on opposite walls are 
combined to assemble the desired physical 4-spinors.

Since these are the first simulations of DWF in dynamical QCD there
are no previous results that would allow an independent check of the
methods and code. Tests using the chiral condensate from the free
field analytical results of \cite{Vranas:1997tj,Vranas:1997da} 
were done in order to check the Dirac operator and inverter. The 
subtraction of Pauli-Villars fields was tested by performing 
simulations with $m_f=1$ and comparing with equivalent results 
from quenched simulations.  Finally, two flavor dynamical simulations 
were done on $2^4$ lattices
and the results were compared with simulations using the overlap
formalism\cite{Narayanan:1993wx,Narayanan:1993ss,Narayanan:1994sk,Narayanan:1995gw} relevant for the DWF action \cite{Furman:1995ky}
for the same parameters. In particular for $\beta=5.6$, $m_f=0.1$,
$m_0=0.9$ the overlap simulation gave
$\langle\psibar\psi\rangle = 1.672(2) \times 10^{-3}$ and average plaquette
$\plaq = 5.765(79) \times 10^{-1}$ while the DWF simulation
with $L_s=18$ gave
$\langle\psibar\psi\rangle = 1.653(33) \times 10^{-3}$ and average plaquette
$\plaq = 5.841(47) \times 10^{-1}$.

All numerical results in this work were obtained from lattices 
of size $L=8$, $N_t=4$ with periodic spatial boundary conditions 
and anti-periodic temporal boundary conditions. The fifth 
direction was set to various values in the range $[8, 40]$, 
the domain wall height $m_0$ was varied in the range 
$[1.15, 2.4]$, the fermion mass was varied in the range 
$[0.02, 0.18]$ and $\beta$ was varied in the range $[4.65, 5.95]$.  
The molecular dynamics trajectory length was set to $\tau = 0.5$ 
and the step size $\delta\tau$ was set to various values in the range 
$[0.0078, 0.02]$ depending on the values of the other parameters.
The CG stopping condition which is defined as the ratio
of the norm of the residual vector over the norm of the source was set 
to $10^{-6}$.  This resulted in an average number of CG iterations 
ranging between 50 and 400 depending on the values of the other parameters. 

The initial configuration was generally chosen to be in the phase 
opposite to that expected for the input parameters creating a very 
visible thermalization process in which the system should be seen to 
evolve into the correct phase.  Typically $100 - 400$ trajectories 
were needed to thermalize the lattice.  The chiral condensate and 
Wilson line were measured in every sweep. The chiral condensate was 
measured using a standard ``one-hit'' stochastic estimator of the 
trace of $D_F^{-1}$ with spin and $s$ coordinates restricted 
according to Eq.~\ref{projection}.  Specifically we evaluated the
quantities:

\begin{eqnarray}
\langle|W|\rangle &=& {1\over 3 L^3} \Big|\sum_{\vec x} 
                            {\rm tr}[\prod_{l\in L(\vec x)} U_l]\Big| \\
\langle \psibar\psi \rangle &=& {1\over 12 L^3 N_t} \Big\{
                   {\rm tr}[\langle s=0|1/D_F| s=L_s-1 \rangle P_{\rm R}] 
                                                        \nonumber \\
      &&\quad\quad+{\rm tr}[\langle s=L_s-1|1/D_F| s=0 \rangle P_{\rm L}]
                   \Big\}.
\label{eq:psibarpsi}
\end{eqnarray}
Here $U_l$ identifies the $SU(3)$ gauge matrix corresponding to the 
link $l$ and the ordered product is taken for all links in the 
time-like line $L(\vec x)$ with spatial coordinate $\vec x$.  The 
somewhat unconventional normalization in Eq.~\ref{eq:psibarpsi} was 
used in our previous work and determines a spin and color average 
which for very large mass $m_f$ approaches $1/m_f$.  (Note, here 
$D_F$ is the single-flavor Dirac operator defined in Eq.~\ref{eq:D_F}.)

\section{Analytical considerations}
\label{sec:analytical}

In this section we summarize some of the analytically determined
properties of domain wall fermions.  These help guide our numerical
investigations, which are done for finite and non-zero values for the
three parameters of domain wall fermions, $L_s$, $m_0$, and $m_f$, as
well as at finite bare coupling $g_0$.

\subsection{$L_s$ dependence}

For numerical simulations, the existence of the chiral limit for domain
wall fermions and the rate of approach to it are of primary
importance.  The computational requirements for domain wall fermions
grow as one power of $L_s$ from the simple increase in the number of
operations.  An additional slight increase in computational cost
for larger $L_s$ comes from the decrease in the total quark mass
due to smaller mixing between the chiral surface states, until
the quark mass is dominated by the input $m_f$.

The axial Ward-Takahashi identities for domain wall fermions are the
same as the continuum, except for an additional term which comes from
the mixing of the left- and right-handed light surface states at the
midpoint of the fifth dimension, $L_s/2$.  At any lattice spacing this
additional term vanishes as $L_s \rightarrow \infty$ for non-singlet
axial symmetries
\cite{Furman:1995ky,Narayanan:1993wx,Narayanan:1993ss,Narayanan:1994sk,Narayanan:1995gw}.
For the singlet axial symmetry, this extra term generates the axial
anomaly.  At strong coupling, the axial currents are conserved for 
$L_s \rightarrow \infty$ but, since the doubler fermions may enter 
the spectrum, these currents may not have the physical significance 
of axial currents\cite{Furman:1995ky}.

For free domain wall fermions, the rate of approach to the chiral limit
can be calculated.  At finite $L_s$ the mixing of the chiral components
is reflected in the fermion mass $m_{\rm eff}$.  For the one
flavor theory this effective mass is \cite{Vranas:1997da}
\begin{equation}
m_{\rm eff} = m_0 (2 - m_0) \left[ m_f + (1-m_0)^{L_s}\right]
  \ \ \ \ \ \ 0 < m_0 < 2,
\label{eq:meff_free}
\end{equation}
$m_{\rm eff}$ has two pieces: one is proportional to the bare mass
$m_f$ and the other expresses the residual mixing between the chiral
modes bound to the domain walls.  Since each bound chiral state decays
exponentially with the distance from its wall, the residual mixing
between them vanishes exponentially with $L_s$, with a decay constant
of $-\ln{|1-m_0|}$.  Notice that when $L_s \rightarrow \infty$, $m_0$
becomes an irrelevant parameter, provided it stays in the range (0,2).

In the free theory, one also finds that fermion states with non-zero
four-momentum decay more slowly with the distance from the wall than
do zero momentum states.   The decay is controlled by the four-momentum
and the value for $m_0$.  Since the lattice momentum $p^\mu_L = p^\mu a$,
where $a$ is the lattice spacing, the slower decay for modes with
non-zero four-momentum is an $O(a^2)$ effect which should vanish in
the continuum limit.  In addition, for a given $m_0$, there is
a critical four-momentum above which the fermions are no longer bound to
the wall, but instead behave like massive, five-dimensional fermions.
Of course, because these fermions are massive, they necessarily preserve 
the theory's four-dimensional chiral symmetry since their propagation 
between the $s=0$ and $s=L_s-1$ walls is exponentially suppressed.

For interacting theories, a simple expectation is for 
Eq.~\ref{eq:meff_free} to be replaced by
\begin{equation}
m_{\rm eff} = Z_m \left[ m_f + c ~e^{ - \alpha L_s}\right].
\label{meff_univ}
\end{equation}
The exponential dependence is seen perturbatively 
\cite{Shamir:1993zy,Aoki:1997xg,Kikukawa:1997tf} and proven to exist 
non-perturbatively, provided the gauge fields satisfy a smoothness 
condition\cite{Hernandez:1998et,Neuberger:1999pz}.  These analytic 
results support the expectation of exponential suppression of 
chiral symmetry breaking effects in the non-perturbative regime.  
However, this behavior may be best established by the sort of 
explicit numerical study reported here.  Generally $\alpha$ should 
depend on $m_0$, allowing one to choose an optimal value for 
simulations at finite $L_s$.  While in the free theory $m_0=1$ 
gives $e^{-\alpha} = 0$, for the interacting theory the variable 
character of fermion propagation in fluctuating background gauge 
fields makes decoupling the walls with a single value for $m_0$ 
unlikely, except at very weak coupling.

Close to the continuum limit, it can be argued that this form for the
effective mass, an input quark mass plus a residual mass $m_{\rm res}$,
should enter all long-distance observables.  However, away from the 
continuum limit or for quantities that cannot be obtained from a low 
energy effective QCD Lagrangian this is not necessarily the case. 
Therefore, different observables may approach their $L_s\rightarrow\infty$ 
limit in different ways, depending on the momentum scales which 
enter the observable, and the corrections to the input quark mass, 
particularly at stronger couplings, may be more complicated.  
In a numerical investigation this has to be kept in mind.  
In this paper only the chiral condensate and pion susceptibility 
are considered.  Work on larger lattices involving measurements 
of many fermionic operators is currently in progress\cite{Columbia_cur}.

Numerical simulations may well be the only way to determine the
dependence of chiral symmetry breaking effects on $L_s$ for
intermediate lattice spacings ($\sim 1$ to 3 GeV).  While for full QCD,
perturbative and non-perturbative arguments support exponential falloff
with $L_s$, for quenched theories, where the lack of damping
from a fermionic determinant can lead to configurations with
unsuppressed small eigenvalues for the fermions, the large $L_s$
behavior is even more in need of determination through simulations.
Some results from quenched QCD simulations have been discussed in
Refs.~\cite{Blum:1997mz,Blum:1997jf,Mawhinney:1998ut,Fleming:1998cc,Kaehler:1998xx,Edwards:1998sh,Fleming:1999eq,Wu:1999cd,AliKhan:1999zn,Edwards:1999bm,Aoki:1999uv,Aoki:2000pc,Chen:1998xw}.

\subsection{$m_0$ dependence}

For free domain wall fermions the number of light flavors is
controlled by the value of $m_0$ \cite{Kaplan:1993sg}. In particular
$m_0 < 0$ corresponds to zero light flavors,
$0 < m_0 < 2$ to one,
$2 < m_0 < 4$ to four, and
$4 < m_0 < 6$ to six light flavors. 
The theory is symmetric under $m_0 \rightarrow 10 - m_0$.

For the interacting theory the values of $m_0$ which distinguish
between different numbers of flavors are changed.  Light fermions first
appear for $m_0 > 0$, the one to four flavor transition occurs for $m_0
> 2$, etc.\ and the theory is still symmetric about $m_0 \rightarrow 10
- m_0$.  This is expected perturbatively and seen numerically
\cite{Mawhinney:1998ut,Chen:1998xw,Vranas:1999pm}.  There is also
some numerical evidence that the transition between different numbers
of flavors is smooth and spread out over a small region of $m_0$
\cite{Mawhinney:1998ut}.  For the interacting theory, keeping $m_0 < 2$
guarantees that a theory with not more than one flavor is being studied.

While $m_0$ is an irrelevant parameter for $L_s \rightarrow \infty$, it
is very important for simulations, not only in controlling the approach
to the chiral limit and the flavor content of the theory, but also for
insuring that light fermions with an average momentum given by the
temperature are still bound to the walls.  For the free theory, the
range of four-momenta carried by states that are bound to the walls
increases as $m_0$ increases from zero, as do the corresponding 
Dirac eigenvalues.  As $m_0$ approaches one, the largest Dirac 
eigenvalues of these ``bound'' states become farther off-shell, 
with values $\approx 1/a$.  As $m_0$ increases above 1, the
number of these off-shell states continues to grow but rather than
their eigenvalues increasing, instead their degeneracy increases beyond
what would be seen for the large momentum states of a free theory.
As $m_0$ increases further and approaches 2, some of these excess, degenerate
states become more nearly on-shell until for $m_0>2$ one has the low-lying
Dirac eigenvalues of a free, four-flavor theory.  Thus, in the free case
a choice of $m_0$ midway between 0 and 2 is best, giving the largest
phase space for physical states bound to the walls, without
adding additional flavors.  Using this behavior as a guide for the 
interacting case, one expects that choosing $m_0$ midway between the 
value where  a single light fermion is bound to the walls and four 
light fermions are bound allows the largest range of four-momentum 
for a single flavor of light quark bound to the walls.

\subsection{Topology}

An important property of the domain wall fermion Dirac operator is the
presence of exact zero modes in the $L_s=\infty$ limit, as can be seen
from the overlap formalism 
\cite{Narayanan:1993wx,Narayanan:1993ss,Narayanan:1994sk,Narayanan:1995gw}.  
These zero modes are related to
the topological charge of the gauge field and as a result an
approximate form of the index theorem is present on the lattice
\cite{Narayanan:1997sa}.  Studies on semiclassical configurations show
the presence of modes which are very close to zero modes even at finite
$L_s$ \cite{Chen:1998ne} and as a result make lattice studies of
anomalous symmetry breaking possible \cite{Vranas:1998vm,Chen:1998xw,Vranas:1999pm}

During simulations, field configurations of different winding number
should show zero mode effects in fermionic observables.  The efficiency
with which the hybrid Monte Carlo can move the system between sectors
of different winding is an important question, as are the long
correlations along the fifth direction which develop for gauge field
configurations where the topology is changing.  These issues have been
studied in numerical simulations of the dynamical Schwinger model
\cite{Vranas:1997da} where the hybrid Monte Carlo performed well
and topology changing occurred.  For this exploratory study of full QCD
thermodynamics, the input quark masses are not small, so the effects of
topology should not be particularly large.

\section{The finite temperature QCD phase transition}
\label{sec:trans}

The previous sections have described the domain wall fermion
formulation and important questions about it that need to be
investigated numerically.  Here we report on simulations of full QCD at
finite temperature with domain wall fermions on $8^3 \times 4$
lattices.  Studying this system allows us to investigate domain wall
fermions for full QCD and look for the presence of chirally broken and
symmetric phases.  The small volume makes scanning over many values for
$L_s$, $m_0$, $m_f$ and $g_0$ possible, laying the foundation for more
realistic simulations on larger volume.

Since the finite temperature transition of QCD is controlled by the
chiral symmetries of the theory (for light quarks), using domain wall
fermions to preserve the full global symmetries of the continuum
should remove one systematic lattice error that is difficult to control. 
However, finite temperature simulations are generally only possible on
relatively coarse lattices ($a^{-1} \sim 700$ MeV for a lattice with
$N_t = 4$), where analytic results about domain wall fermions are
lacking.  The light chiral modes of domain wall fermions at weak
coupling must exist at $a^{-1} \sim 700$ MeV, in the full
non-perturbative gauge field backgrounds, for thermodynamic simulations
to be possible.  If it is found that chiral modes exist on coarse
lattices, the size of the $m_{\rm res}(L_s, \beta)$ and its dependence
on $L_s$ and $\beta$ must be investigated.  (As already mentioned,
$m_{\rm res}$ is only a sensible quantity for low-energy
observables and it must be demonstrated that various determinations
of it are consistent.  In this section we refer to $m_{\rm res}$,
without specifying precisely how it may be determined, as a generic
indicator of the mixing between the chiral modes.)

\subsection{Locating the transition}

Locating the phase transition in full QCD requires scanning values for
four parameters ($m_0$, $L_s$, $m_f$, and $\beta$).  Without any
knowledge of the location of the transition, or if it even exists for
domain wall fermions, choosing parameters for initial simulations is
difficult.  For staggered fermions, the critical coupling for the
finite temperature phase transition for 2 flavors on an $N_t = 4 $
lattice is $\beta_c = 5.265$ for $m = 0.01$ and $\beta_c = 5.291$ for
$m = 0.025$ \cite{Vaccarino:1991jy}.  Since staggered and domain wall
fermions both have their chiral limit at zero quark mass, the light
quarks have the largest effect in the location of $\beta_c$ and both
theories have the same number of light flavors, we used the staggered
values as a rough guide.

Our first simulations with domain wall fermions were done at $\beta =
5.0$ and $\beta = 5.4$, with the hope that these would be above and
below the transition region.  $L_s = 8$ and $m_f = 0.1$ were chosen to
keep the computational difficulty modest.  We worked with $m_0 = 1.65$,
since for quenched simulations this choice gave a reasonable falloff
between the walls at $\beta = 5.7$ and for quenched QCD, $\beta = 5.7$
is close to $\beta_c$ for an $N_t = 4 $ lattice.  

\begin{figure}
\epsfxsize=\hsize
\epsfbox{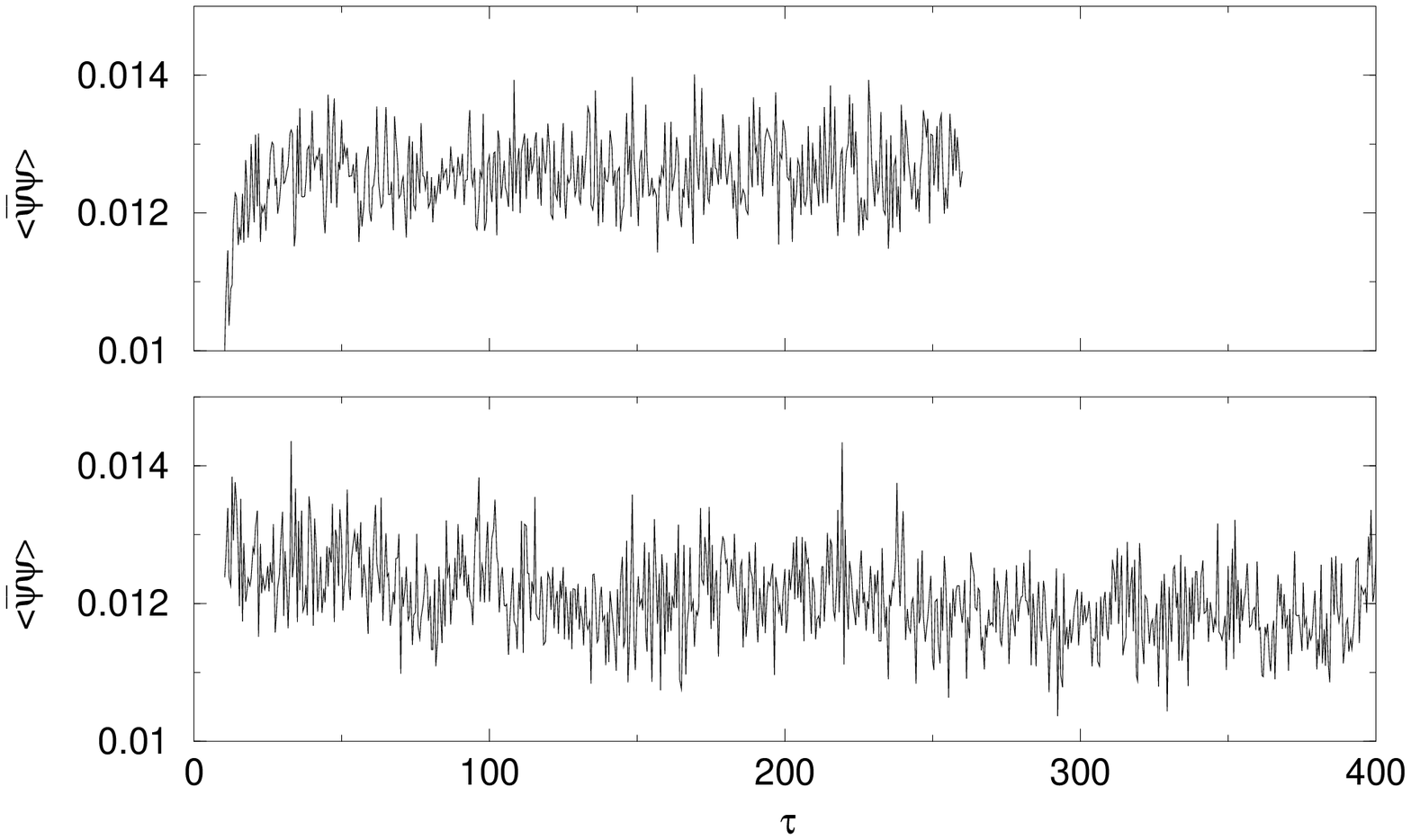}
\caption{ The simulation ``time history" of $\langle \bar{\psi} \psi
\rangle$ for $\beta = 5.0$ (upper graph) and $\beta = 5.4$ lower graph
for $m_f = 0.1$, $m_0 = 1.65$ and $L_s = 8$.}
\label{fig:b5_0_b5_4_ls8_M1_65_evol}
\end{figure}

Although with this choice of $m_0$, the $\beta$ range being examined
($5.0\le\beta\le5.4$) lies below the chiral transition,
we describe this point first since it demonstrates our
very first efforts in charting this parameter space and the difficulties
we encountered.  The evolution of $\langle \bar{\psi} \psi \rangle$ for 
$m_f = 0.1$ and $\beta = 5.0$ is shown in the upper panel of 
Figure~\ref{fig:b5_0_b5_4_ls8_M1_65_evol} and the lower panel is for 
$\beta = 5.4$.  The hybrid Monte Carlo was run with a step size of 
$\delta\tau = 0.025$ and 20 steps per trajectory, giving an acceptance of 
66\% for $\beta = 5.0$ and 70\% for $\beta = 5.4$.  The evolution appears 
quite generic and the simulation presented no difficulty to the
hybrid Monte Carlo.  For $\beta = 5.0$ the Wilson line expectation
value was 0.0223(15) and for $\beta = 5.4$ it was 0.0466(41).  Both
these values are small and indicate that both $\beta$ values correspond
to the confined phase.

The chiral condensate $\langle \bar{\psi} \psi \rangle$ was also 
measured for a variety of valence masses.  In quenched QCD at 
zero temperature, extrapolations of $\langle \bar{\psi} \psi \rangle$ 
to $m_f = 0$ using quark masses from $m_f = 0.02$ to $m_f = 0.1$ were 
used to see that chiral modes existed for a particular 
$m_0$\cite{Mawhinney:1998ut}.   The limit $\langle \bar{\psi}
\psi \rangle(m_f \rightarrow 0 )$ could only be non-zero if light
chiral modes were present, provided $L_s$ is large enough that the
residual mixing is unimportant.  (For the current finite temperature
case, $\langle \bar{\psi} \psi \rangle(m_f \rightarrow 0)$ can be zero
either from the absence of chiral modes or because the system is
in the symmetry-restored phase.)

\begin{figure}
\epsfxsize=\hsize
\epsfbox{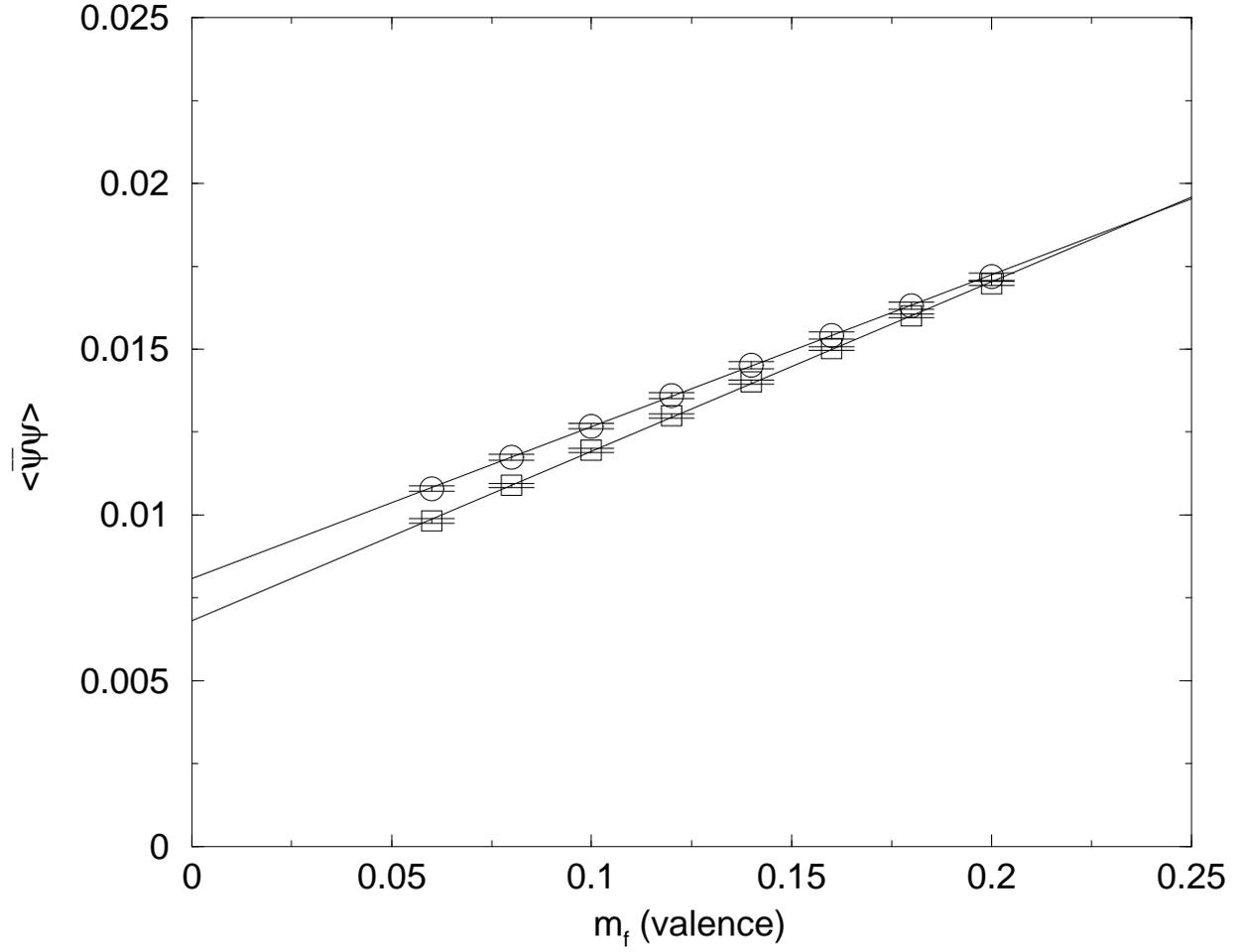}
\caption{ Valence extrapolations of $\langle \bar{\psi} \psi \rangle$
for $m_f = 0.1$, $m_0 = 1.65$ and $L_s = 8$.  The circles are for
$\beta = 5.0$, the squares for $\beta = 5.4$.}
\label{fig:ls8_M1_65_quench}
\end{figure}

Figure \ref{fig:ls8_M1_65_quench} shows that $\langle \bar{\psi} \psi
\rangle$ extrapolates to a non-zero value for both $\beta = 5.0$ and
$\beta = 5.4$ and this value is not very sensitive to $\beta$.  The
values for the Wilson line indicate both $\beta$ values are in the
confined phase, so the $\langle \bar{\psi} \psi \rangle$ results
show that light chiral modes are present with an unknown residual
mixing.  The insensitivity to $\beta$ is an interesting feature.

Next, instead of scanning larger values of $\beta$, we decided to
change $m_0$ from 1.65 to 1.9, keeping all other parameters identical.
(This reflects our initial search path in parameter space and does not
imply the absence of a transition at $m_0=1.65$ and $L_s=8$.)
The acceptance is 59\% for $\beta = 5.0$ and 71\% for $\beta = 5.4$
The Wilson line for $\beta = 5.0$ is 0.030(2), while for 
$\beta = 5.4$ it is 0.202(5), indicating that $\beta = 5.4$ 
is likely deconfined.  The evolutions show a very different 
behavior for the condensate evaluated at the dynamical quark
mass.  The value at $\beta=5.0$ has increased, part of which likely
reflects the change with $m_0$ in the overlap between the
five-dimensional light modes and the surfaces at $s = 0$ and $L_s -
1$.  The $\beta=5.4$ values are much smaller, consistent with the
deconfined phase.

\begin{figure}
\epsfxsize=\hsize
\epsfbox{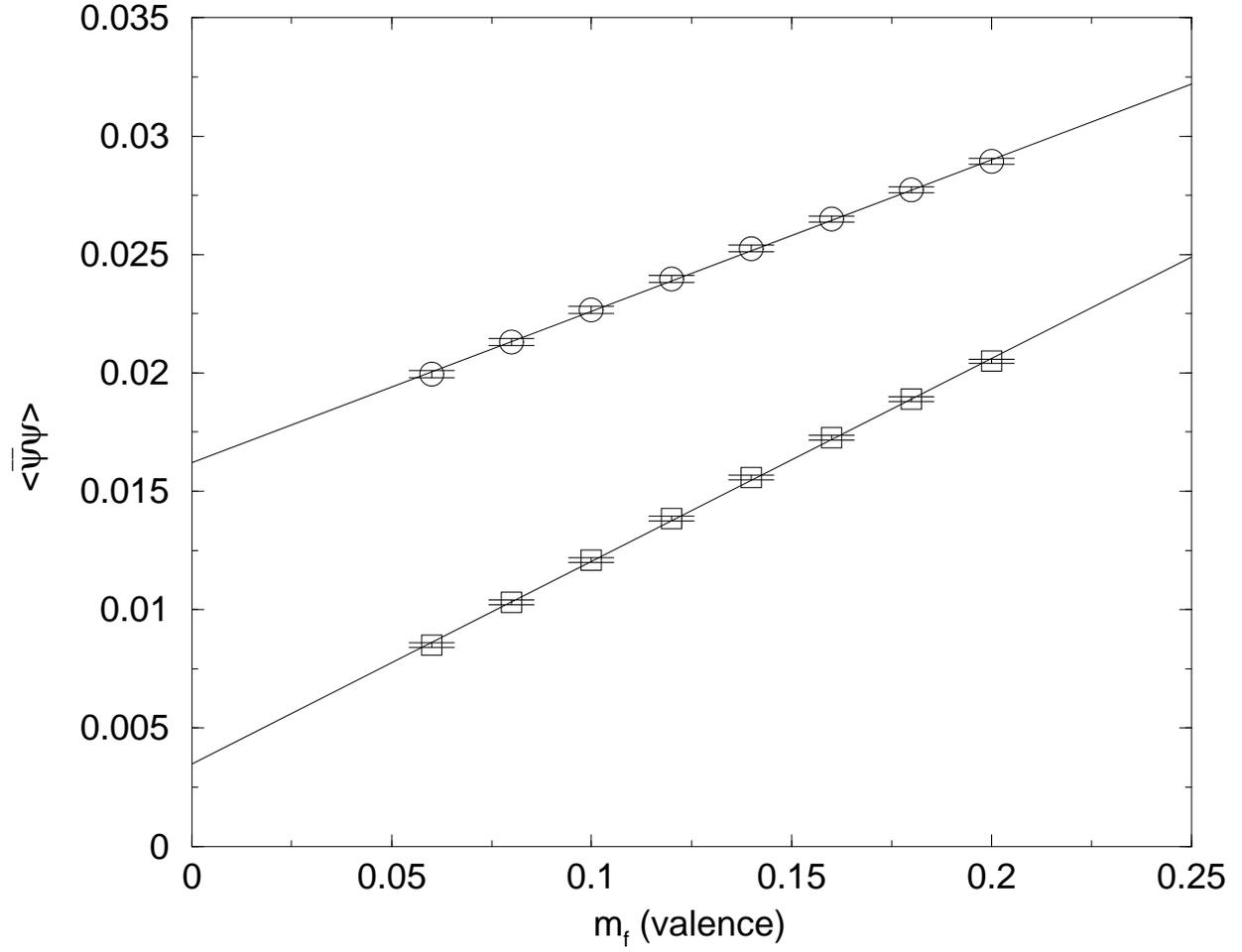}
\caption{ Valence extrapolations of $\langle \bar{\psi} \psi \rangle$
for $m_f = 0.1$, $m_0 = 1.90$ and $L_s = 8$.  The circles are results
for $\beta = 5.0$, the squares are for $\beta = 5.4$.}
\label{fig:ls8_M1_90_quench}
\end{figure}

Figure \ref{fig:ls8_M1_90_quench} shows the valence quark extrapolation.
The small value of $\langle \bar{\psi} \psi \rangle(m_f \rightarrow 0)$
suggests the restoration of chiral symmetry.  Of course, there is a 
possibility that this small value might instead be caused by the loss 
of chiral modes.  However, this is unlikely because we have seen that 
chiral modes do exist for $\beta=5.0$ and one expects that at the weaker 
$\beta=5.4$ coupling these chiral modes should be even more numerous.  
Therefore, we have preliminary evidence for two phases of full QCD 
with dynamical domain wall fermions.

\begin{figure}
\epsfxsize=\hsize
\epsfbox{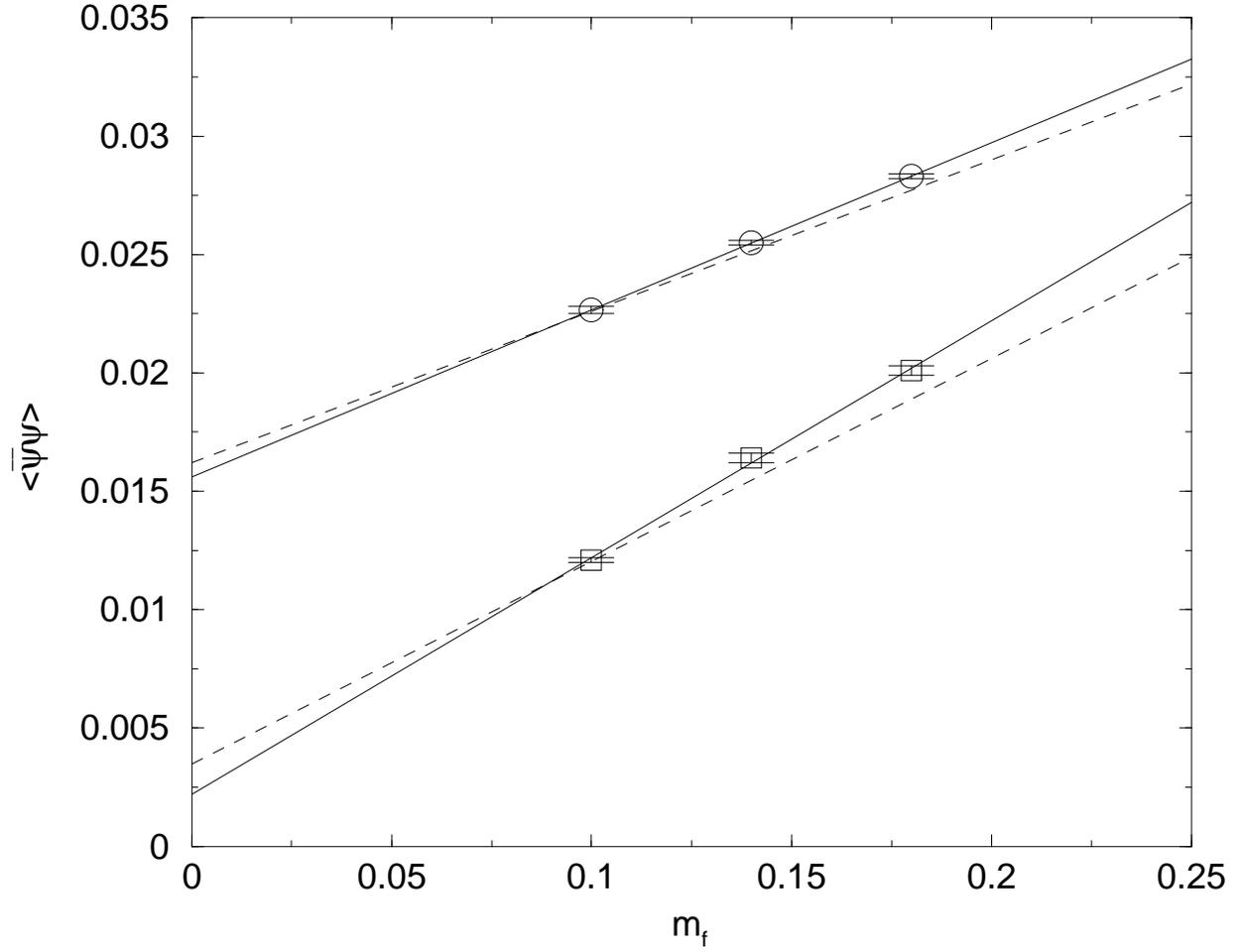}
\caption{ Full QCD extrapolations of $\langle\bar{\psi}\psi\rangle$ for
$m_0 = 1.90$ and $L_s = 8$.  The circles are results for $\beta = 5.0$,
the squares are for $\beta = 5.4$.  The dashed line is the fit to the
quenched data given in Figure \ref{fig:ls8_M1_90_quench} while
the solid line is a fit to the dynamical values for $\langle\bar{\psi}
\psi\rangle$.}
\label{fig:ls8_M1_90_full}
\end{figure}

To solidify the evidence for two different phases of QCD with domain
wall fermions further simulations for $L_s = 8$ and $m_0 = 1.9$ were
done with dynamical quark masses of 0.14 and 0.18.  These points are
shown in Figure \ref{fig:ls8_M1_90_full}.  The dashed line is the fit to
the quenched extrapolation shown in Figure \ref{fig:ls8_M1_90_quench}.
There is not a large difference between the two extrapolations,
although both full QCD extrapolations fall below the quenched
extrapolations, indicating some suppression of small eigenvalues
through the presence of the fermion determinant.  In the next section,
we study the dynamical mass extrapolation of $\langle \bar{\psi} \psi
\rangle $ for larger values of $L_s$ to see if the non-zero value for
$\langle \bar{\psi} \psi \rangle (m_f \rightarrow 0)$ decreases with
increasing $L_s$.

\begin{figure}
\epsfxsize=\hsize
\epsfbox{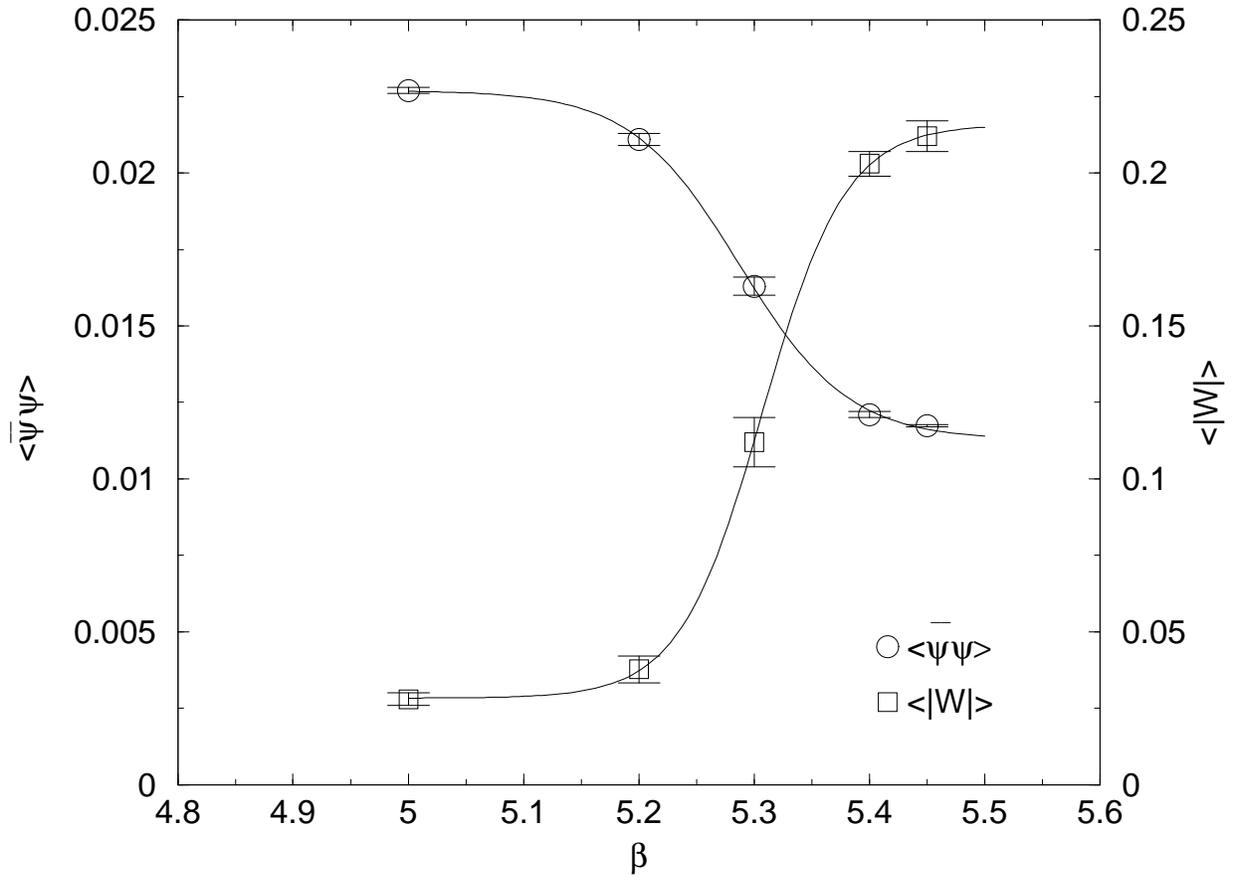}
\caption{ Full QCD values for $\langle \bar{\psi} \psi \rangle$ for
$m_0 = 1.90$, $L_s = 8$ and $m_f = 0.1$ for different values of
$\beta$.}
\label{fig:ls8_betac}
\end{figure}

Additional simulations with $m_f = 0.1$, $L_s = 8$ and $m_0 = 1.9$ were
done for $\beta = $ 5.2, 5.3 and 5.45, which produced the data for
$\langle \bar{\psi} \psi \rangle$ and the Wilson line shown in Figure
\ref{fig:ls8_betac}.  Crossover behavior is seen for both observables
further supporting the identification of both a chirally broken and a
chirally restored phase.  These
simulations are at a small value of $L_s$, so the contribution of
$m_{\rm res}$ to the effective quark mass may be large.  Since $m_{\rm
res}(\beta, L_s)$ is likely varying across the transition region, due
to the change in $\beta$, the shape of the curves is expected to
reflect this varying effective quark mass.

\subsection{$L_s$ dependence in the two phases}

\begin{figure}
\epsfxsize=\hsize
\epsfbox{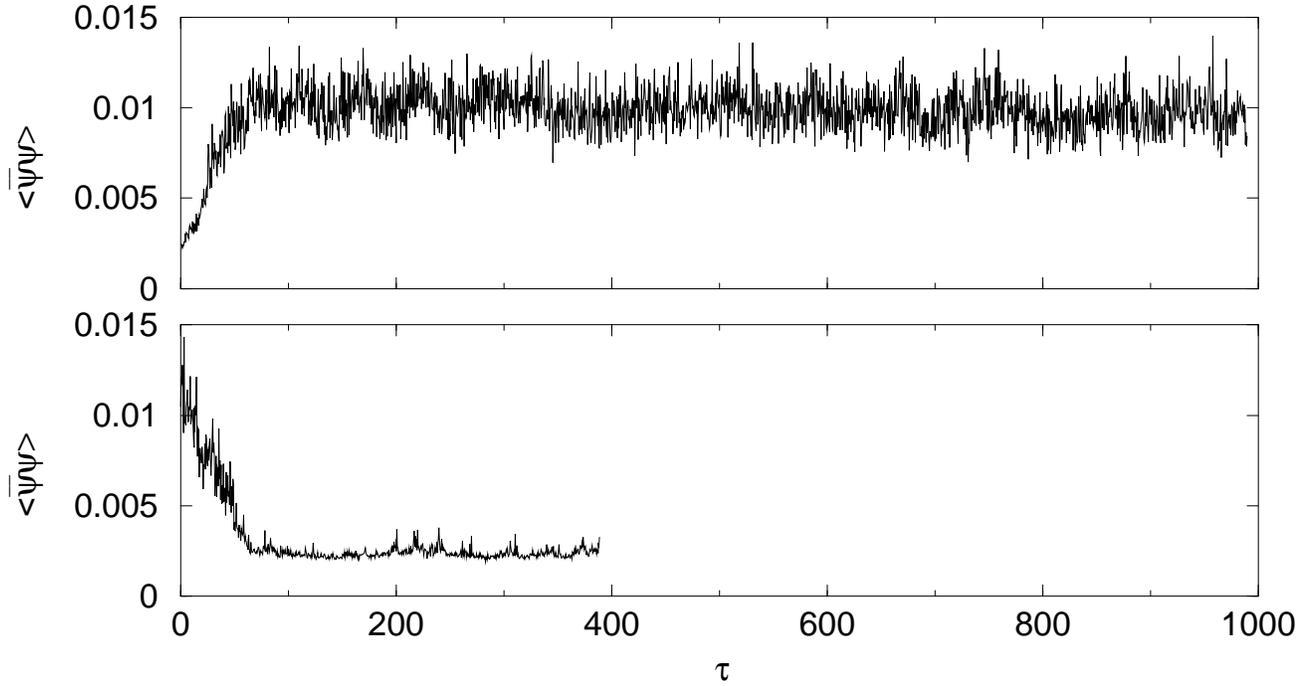}
\caption{ The simulation ``time history" of $\langle \bar{\psi} \psi
\rangle $ for $\beta = 5.2$ (upper graph) and $\beta = 5.45$ (lower
graph) for $m_f = 0.02$, $m_0 = 1.9$ and $L_s = 16$.  The initial
configuration was chosen in the opposite phase, i.e.\ ordered for
$\beta = 5.2$ and disordered for $\beta = 5.45$.}
\label{fig:b5_2_b5_45_evol}
\end{figure}

With this evidence for two phases, we turned to exploring the $L_s$
dependence in each phase.  For the confined phase, we chose $\beta =
5.2$ to be at weaker coupling while still in this phase and in the
deconfined phase we chose $\beta = 5.45$, to be farther from the
transition.  Keeping $m_0 = 1.90$, simulations were done for many
values of $L_s$ and the dynamical quark mass, $m_f$.  Table
\ref{tab:8nt4_b5.2_h1.9} gives the parameters for $\beta = 5.2$ and
Table \ref{tab:8nt4_b5.45_h1.9} gives them for $\beta = 5.45$.  A plot
of the evolution of $\langle \bar{\psi} \psi \rangle$ for $\beta =
5.20$ and 5.45 is shown in Figure \ref{fig:b5_2_b5_45_evol} for $m_f =
0.02$ and $L_s = 16$.  With a step size of $\delta\tau = 1/64$ the
acceptance was 90\%.  Once again there is no evidence for difficulty in
the hybrid Monte Carlo evolution of this system.

\begin{figure}
\epsfxsize=\hsize
\epsfbox{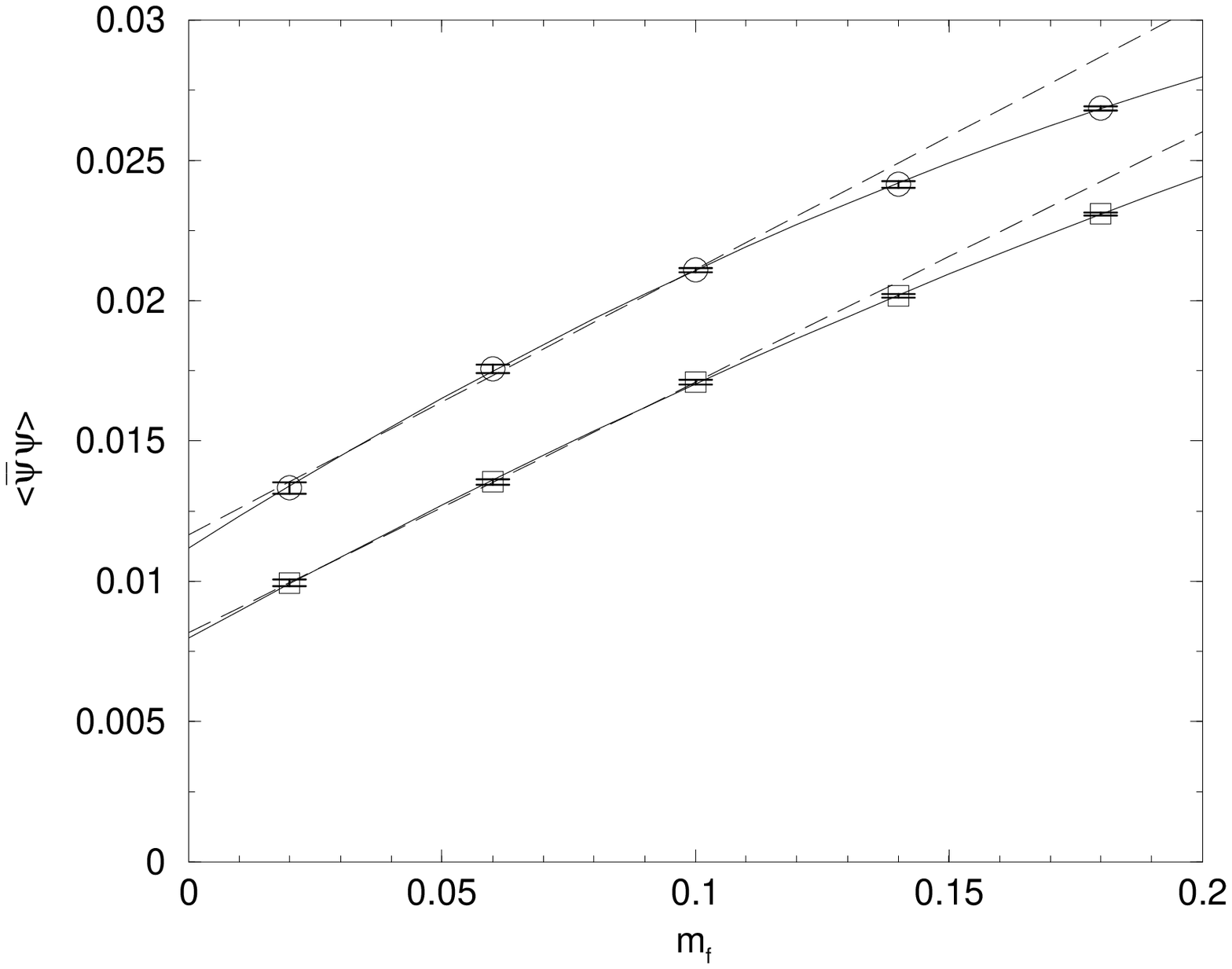}
\caption{ Full QCD values for $\langle \bar{\psi} \psi \rangle$ for
$m_0 = 1.90$ and $\beta = 5.2$ versus $m_f$.  The circles are for
$L_s = 8$ and the squares for 16.}
\label{fig:b5_2_pbp_vs_mf}
\end{figure}

Figure \ref{fig:b5_2_pbp_vs_mf} shows results for $\langle \bar{\psi}
\psi \rangle$ at $\beta = 5.2$ plotted versus $m_f$ for $L_s = 8$ and
16.  The dashed lines are linear fits to the lowest three values for
$m_f$ while the solid lines are quadratic fits to all values of $m_f$.
The fits for $L_s = 8$ are
\begin{eqnarray}
  \langle \bar{\psi} \psi \rangle & = & 0.0117(2) + 0.095(2) m_f \\
  \langle \bar{\psi} \psi \rangle & = & 0.0112(3) + 0.114(5) m_f
    -0.15(2) m_f^2
\end{eqnarray}
with $N_{\rm dof} = 1$ and 2 and $\chi^2/N_{\rm dof} = 3.7$ and 0.4,
respectively.  The fits for $L_s = 16$ are
\begin{eqnarray}
  \langle \bar{\psi} \psi \rangle & = & 0.0082(1) + 0.089(2) m_f \\
  \langle \bar{\psi} \psi \rangle & = & 0.0080(2) + 0.099(3) m_f
    -0.08(1) m_f^2 
\end{eqnarray}
with $N_{\rm dof} = 1$ and 2 and $\chi^2/N_{\rm dof} = 0.03$ and 0.5,
respectively.  The results shows a strong $L_s$ dependence to which
we now turn.

\begin{figure}
\epsfxsize=\hsize
\epsfbox{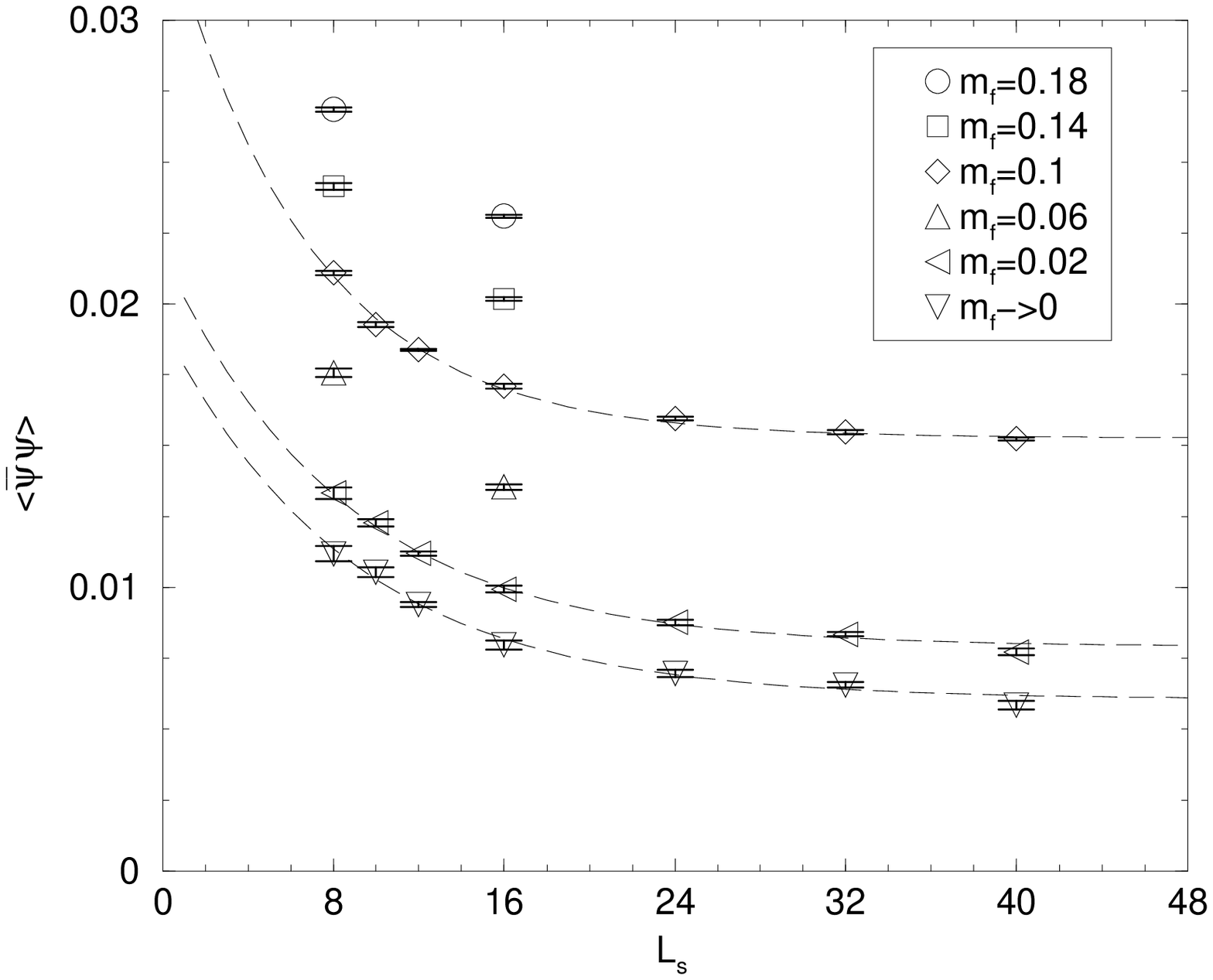}
\caption{ Full QCD values for $\langle \bar{\psi} \psi \rangle$ for
$m_0 = 1.90$ and $\beta = 5.2$ plotted versus $L_s$ for different
values of $m_f$.  The curves are fits of the form
$c_0 + c_1 \exp( -\alpha L_s)$.}
\label{fig:b5_2_pbp_vs_mf_ls}
\end{figure}

Figure \ref{fig:b5_2_pbp_vs_mf_ls} shows $\langle \bar{\psi} \psi
\rangle$ for $\beta = 5.2$ plotted versus $L_s$ for a variety of values
of $m_f$.  The curves are fits to the form $c_0 + c_1 \exp( -\alpha
L_s)$ for $L_s = 8$ to 40.  The fit parameters are
\begin{eqnarray}
  \langle \bar{\psi} \psi \rangle & = & 0.01527(4) + 0.0188(8)
    \exp( -0.149(5) L_s ) \;\;\;  m_f = 0.1 \\
  \langle \bar{\psi} \psi \rangle & = & 0.00779(8) + 0.014(1)
    \exp( -0.116(8) L_s ) \;\;\;\;\;  m_f = 0.02 \\
  \langle \bar{\psi} \psi \rangle & = & 0.0059(1) + 0.014(1)
    \exp( -0.11(1) L_s ) \;\;\;\;\;\;\;  m_f \rightarrow 0.0
\end{eqnarray}
All fits have $N_{\rm dof} = 4$ and give $\chi^2/N_{\rm dof} = 5.1$,
5.6 and 6.6, respectively.  The $m_f \rightarrow 0$ points are first
found by extrapolating to $m_f = 0$ at fixed $L_s$ and then fitting
these values versus $L_s$.  Although the values for $\chi^2$ are
somewhat large, the data is well fit by a function with exponential
dependence on $L_s$.  (Note these somewhat large $\chi^2$ values can be
caused by underestimates of the errors which may result if our Monte
Carlo evolutions are not sufficiently long to allow proper control 
the long-time autocorrelations.)

\begin{figure}
\epsfxsize=\hsize
\epsfbox{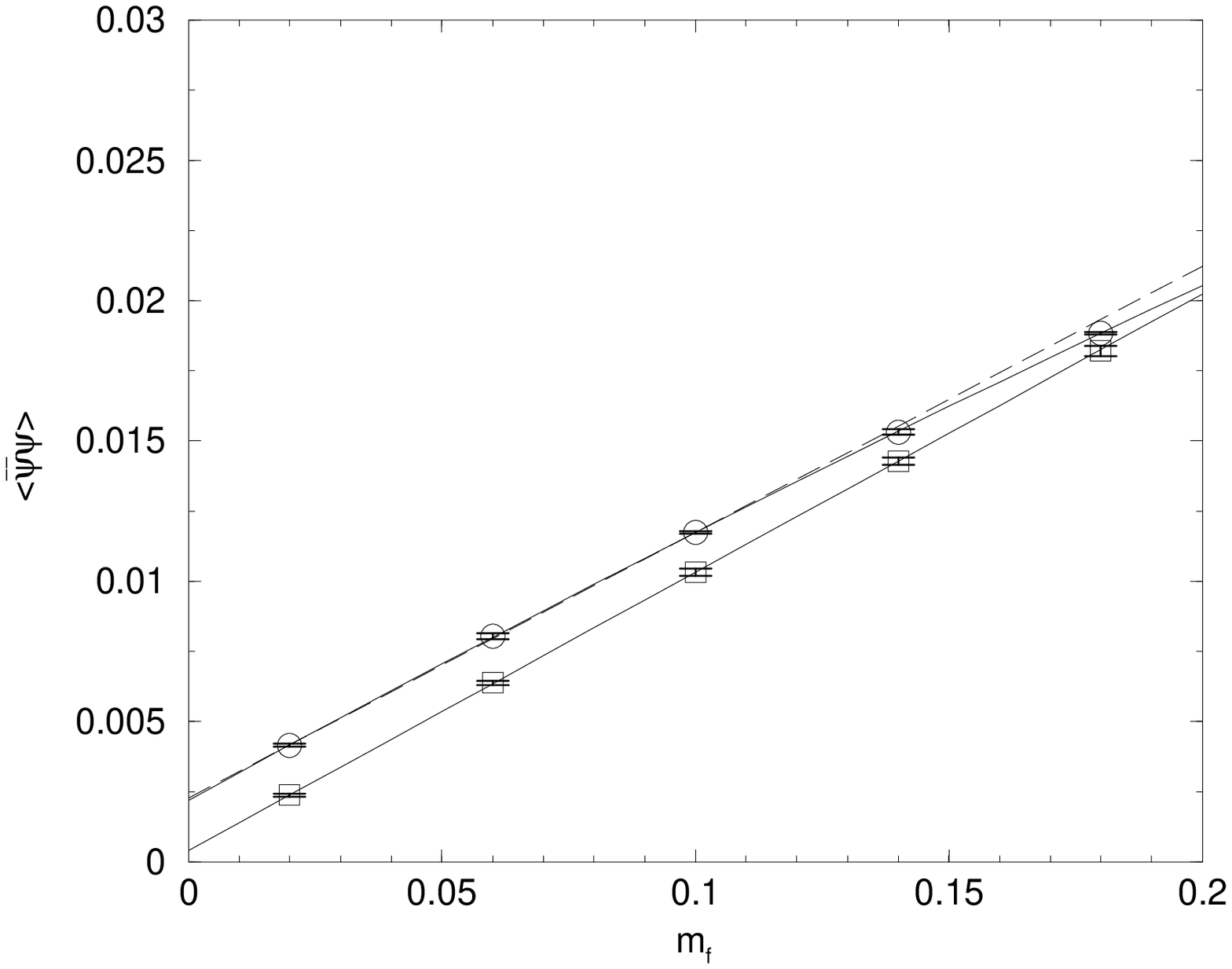}
\caption{ Full QCD values for $\langle \bar{\psi} \psi \rangle$ for
$m_0 = 1.90$ and $\beta = 5.45$ versus $m_f$.  The circles are for
$L_s = 8$, the squares for 16.}
\label{fig:b5_45_pbp_vs_mf}
\end{figure}

Similar results have been obtained for $\beta = 5.45$.  Figure
\ref{fig:b5_45_pbp_vs_mf} shows the results for $\langle \bar{\psi}
\psi \rangle$ for $\beta = 5.45$ for $L_s = 8$ and 16. ($L_s = 24$ and
32 results are tabulated below.)  Again, the dashed lines are linear
fits to the lowest three values for $m_f$ while the solid lines are
quadratic fits to all values of $m_f$.  The fits for $L_s = 8$ are
\begin{eqnarray}
  \langle \bar{\psi} \psi \rangle & = & 0.00227(7) + 0.095(1) m_f \\
  \langle \bar{\psi} \psi \rangle & = & 0.00219(9) + 0.099(2) m_f
    -0.037(9) m_f^2
\end{eqnarray}
with $N_{\rm dof} = 1$ and 2 and $\chi^2/N_{\rm dof} = 0.6$ and 0.1,
respectively.  The fits for $L_s = 16$ are
\begin{eqnarray}
  \langle \bar{\psi} \psi \rangle & = & 0.00039(8) + 0.100(2) m_f \\
  \langle \bar{\psi} \psi \rangle & = & 0.00040(6) + 0.100(3) m_f
    -0.01(2) m_f^2 
\end{eqnarray}
with $N_{\rm dof} = 1$ and 2 and $\chi^2/N_{\rm dof} = 0.09$ and 0.02,
respectively.  Linear fits for the larger values of $L_s$ give
\begin{eqnarray}
  \langle \bar{\psi} \psi \rangle & = & 0.00016(8) + 0.100(2) m_f
    \;\;\; L_s = 24 \\
  \langle \bar{\psi} \psi \rangle & = & 0.00006(6) + 0.099(1) m_f
    \;\;\; L_s = 32
\end{eqnarray}
with $N_{\rm dof} = 1$ for both $L_s$ and $\chi^2/N_{\rm dof} = 0.01$
and 7.1, respectively.  We see that with increasing $L_s$, the
extrapolated value for the condensate at $m_f = 0$ decreases
steadily.

\begin{figure}
\epsfxsize=\hsize
\epsfbox{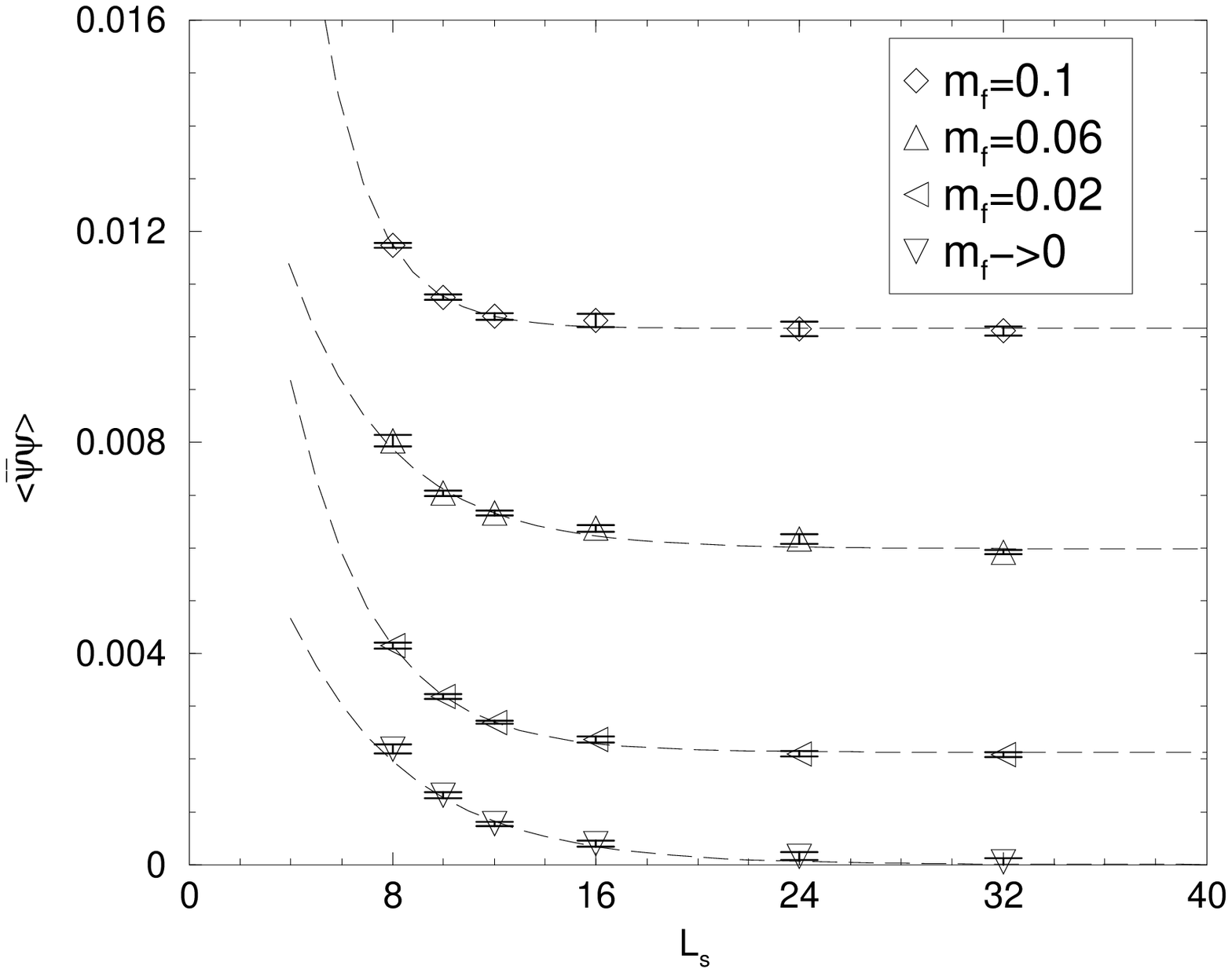}
\caption{ Full QCD values for $\langle \bar{\psi} \psi \rangle$ for
$m_0 = 1.90$ and $\beta = 5.45$ plotted versus $L_s$ for different
values of $m_f$.  The curves are fits of the form
$c_0 + c_1 \exp( -\alpha L_s)$.}
\label{fig:b5_45_pbp_vs_mf_ls}
\end{figure}

Figure \ref{fig:b5_45_pbp_vs_mf_ls} shows $\langle \bar{\psi} \psi
\rangle$ for $\beta = 5.45$ plotted versus $L_s$ for a variety of values
of $m_f$.  The curves are fits to the form $c_0 + c_1 \exp( -\alpha
L_s)$ for $L_s = 8$ to 32.  The fit parameters are
\begin{eqnarray}
  \langle \bar{\psi} \psi \rangle & = & 0.0102(1) + 0.08(3)
    \exp( -0.48(6) L_s ) \;\;\;  m_f = 0.1 \\
  \langle \bar{\psi} \psi \rangle & = & 0.00599(4) + 0.015(3)
    \exp( -0.26(2) L_s ) \;\;\;\;\;  m_f = 0.06 \\
  \langle \bar{\psi} \psi \rangle & = & 0.00213(4) + 0.025(4)
    \exp( -0.31(2) L_s ) \;\;\;\;\;  m_f = 0.02 \\
  \langle \bar{\psi} \psi \rangle & = & 0.00010(5) + 0.019(3)
    \exp( -0.27(2) L_s ) \;\;\;\;\;\;\;  m_f \rightarrow 0.0
\end{eqnarray}
All fits have $N_{\rm dof} = 3$ and give $\chi^2/N_{\rm dof} = 0.4$,
4.8, 1.1 and 0.8, respectively.  Here again the data strongly support
exponential suppression of mixing between the walls for
$\langle \bar{\psi} \psi \rangle$.

For both the confined and deconfined cases, we see $\langle \bar{\psi}
\psi \rangle$ exponentially approaching a limiting value for large
$L_s$ (which is zero in the deconfined case).  At the stronger coupling
of the confined phase, the decay constant is $\sim 1/10$, while in the
deconfined phase it is $\sim 1/4$.  One expects faster decay at weak
coupling, but at present we do not know whether the different phases
also play a role in the decay constant.

\subsection{Studying the $m_0$ dependence of the transition}

The parameter $m_0$ is relevant at finite lattice spacing, since it
controls not only when there is a single light fermion bound to the
domain walls but also the maximum momentum this fermion can have while
still being bound.  It is expected that this parameter will not have to
be fine-tuned for domain wall fermions to work correctly, but care 
in choosing a value is necessary to get the correct number
of light species and the maximum allowable phase space for light 
fermions in the thermal ensemble.

We have studied the characteristics of the transition region
by choosing $m_f = 0.1$, $L_s = 12$ and simulating for
values of $\beta$ near the phase transition for $m_0 = 1.15$,
1.4, 1.65, 1.8, 1.9, 2.0, 2.15 and 2.4.  Tables 
\ref{tab:8nt4_h1.15_beta_crit}, \ref{tab:8nt4_h1.4_beta_crit},
\ref{tab:8nt4_h1.65_beta_crit}, \ref{tab:8nt4_h1.8_beta_crit},
\ref{tab:8nt4_h1.9_beta_crit}, \ref{tab:8nt4_h2.0_beta_crit},
\ref{tab:8nt4_h2.15_beta_crit} and \ref{tab:8nt4_h2.4_beta_crit}
contain simulation parameters and results.  For parameters
where a deconfined thermal state was expected, the initial
lattice was disordered, while an initial ordered lattice was used where
a confined state was expected.

\begin{figure}
\epsfxsize=\hsize
\epsfbox{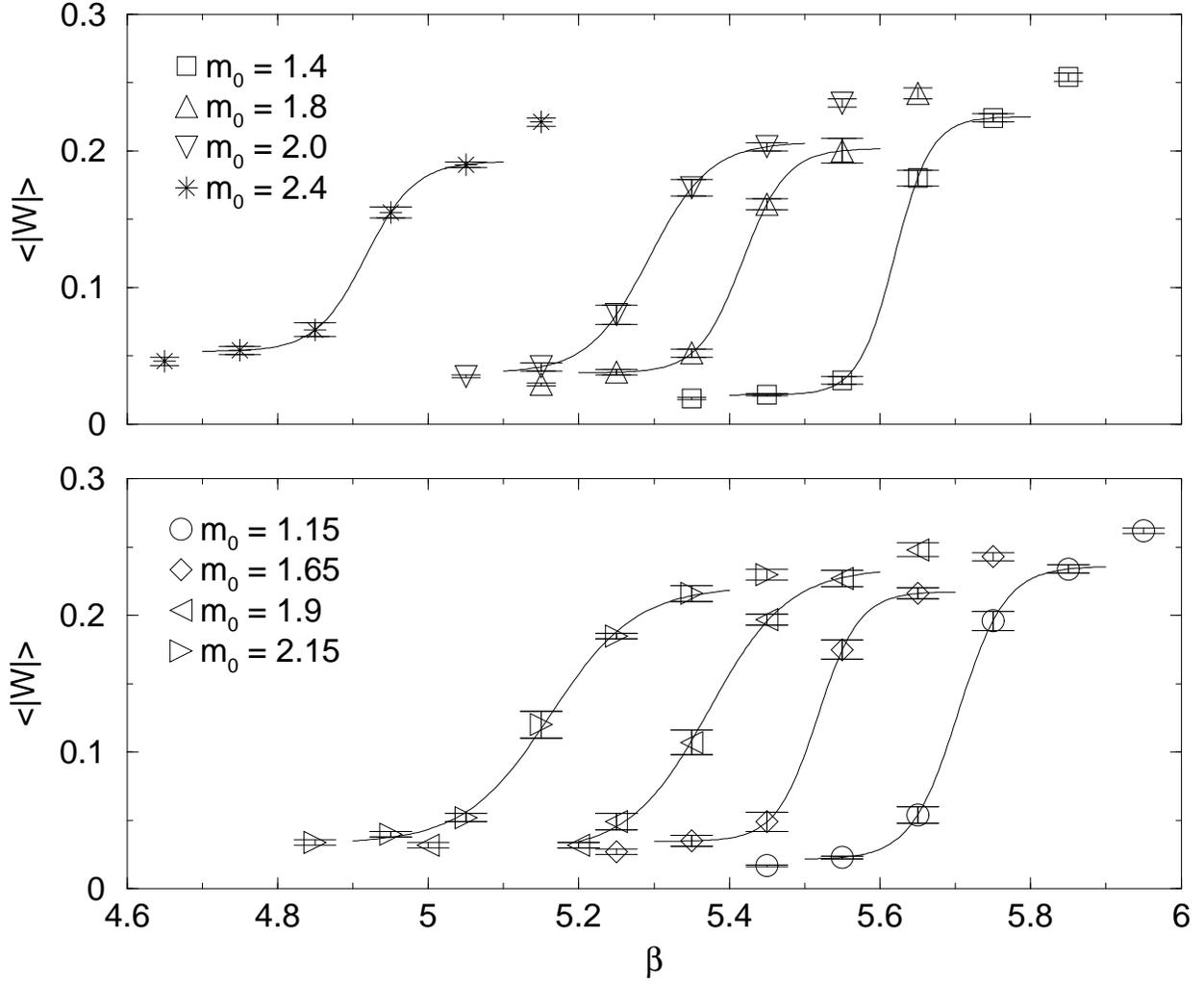}
\caption{ Full QCD results for the Wilson line for $m_f = 0.1$ and
$L_s = 12$ for different values of $m_0$ and $\beta$ 
near the transition region.}
\label{fig:wline_vs_m0}
\end{figure}

Figure \ref{fig:wline_vs_m0} shows the expectation value
of the magnitude of the Wilson line $\langle|W|\rangle$ for these runs.  A rapid
crossover is seen for all values of $m_0$.  The lines are the result
of fitting the four points nearest the transition (five points where
we have a point close to the transition) to the function
\begin{equation}
f(x) = c_0 ( c_1 + \tanh{ [ c_2 (x - \beta_c)] } ).
\label{eq:tanh_fit}
\end{equation}
This is a phenomenologically useful form for determining the point
of maximum slope for the Wilson line.  The points far from the
transition are not included in these fits, since this phenomenological
function poorly represents the data there.

\begin{figure}
\epsfxsize=\hsize
\epsfbox{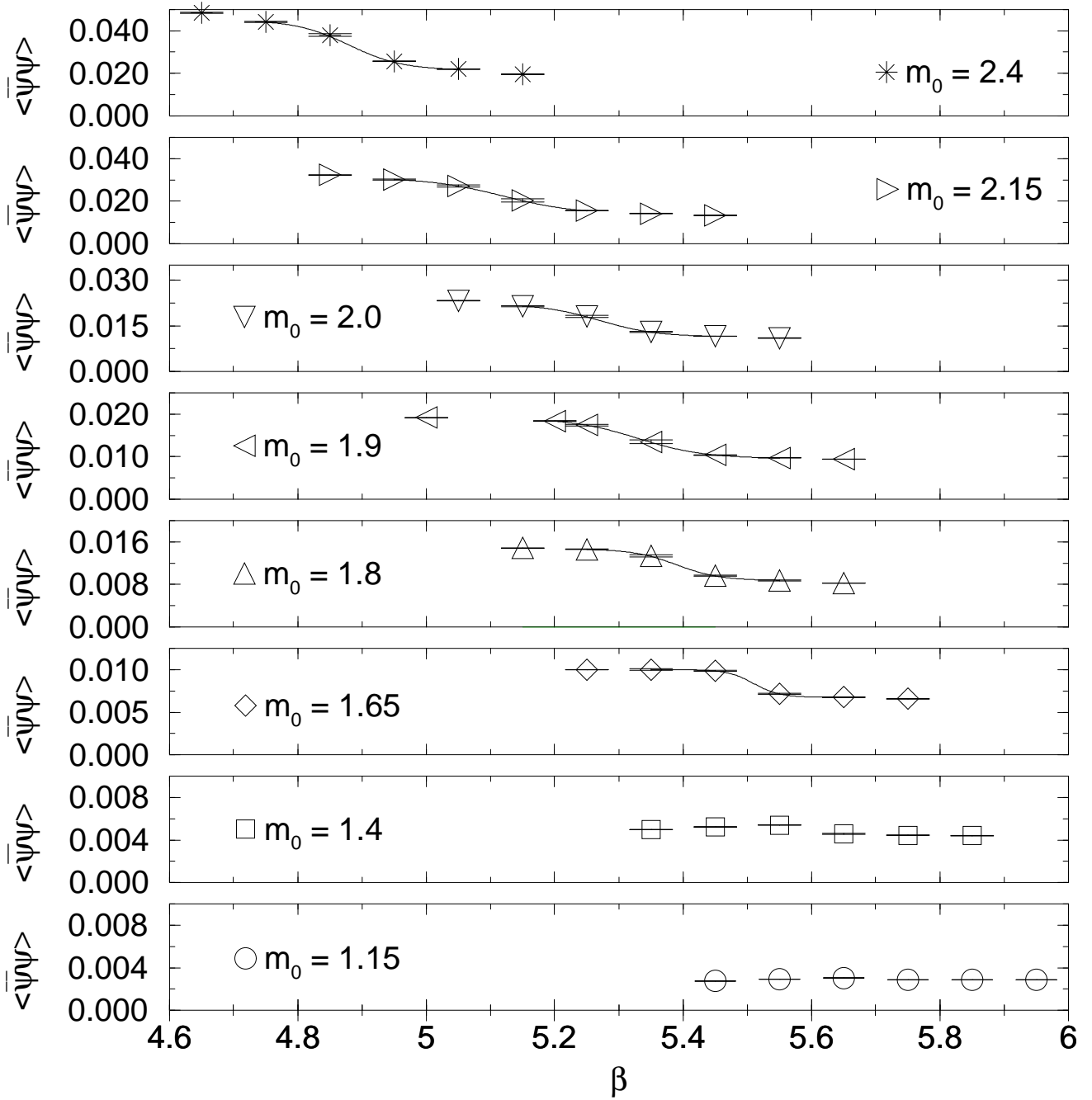}
\caption{ Full QCD results for $\langle\psibar\psi\rangle$ for $m_f =
0.1$ and $L_s = 12$ for different values of $m_0$ and $\beta$ near
the transition region.  Note that the vertical scale decreases
by a factor of five as $m_0$ decreases from 2.4 to 1.15.   This is 
needed to follow the large decrease in the scale of 
$\langle\psibar\psi\rangle$ which results from a combination of 
the decreasing lattice spacing that follows from increasing $\beta$ 
and the diminishing overlap of the light fermion states with the walls.}
\label{fig:pbp_vs_m0}
\end{figure}

Figure \ref{fig:pbp_vs_m0} shows similar results for $\langle
\bar{\psi} \psi \rangle$ with the lines being a fit to 
Eq.~\ref{eq:tanh_fit}.  For $m_0 = 1.15$ and 1.4, the $\langle \bar{\psi}
\psi \rangle$ data do not allow even a rough determination of
$\beta_c$.  For small enough $m_0$, the light chiral modes should not
exist and we have evidence for that at $m_0 = 1.15$.  The value for
$\langle \bar{\psi} \psi \rangle$ is very small and shows little change
even when the Wilson line shows evidence for the transition.  In
addition, the Wilson lines indicate the transition is very close to
the value of 5.6925 for quenched QCD on a $24^3 \times 4$ lattice
\cite{Brown:1988qe} supporting the conclusion that light fermion modes
are not present in the simulations.  The effects of the heavy modes 
are apparently quite well canceled by the Pauli-Villars fields.

\begin{figure}
\epsfxsize=\hsize
\epsfbox{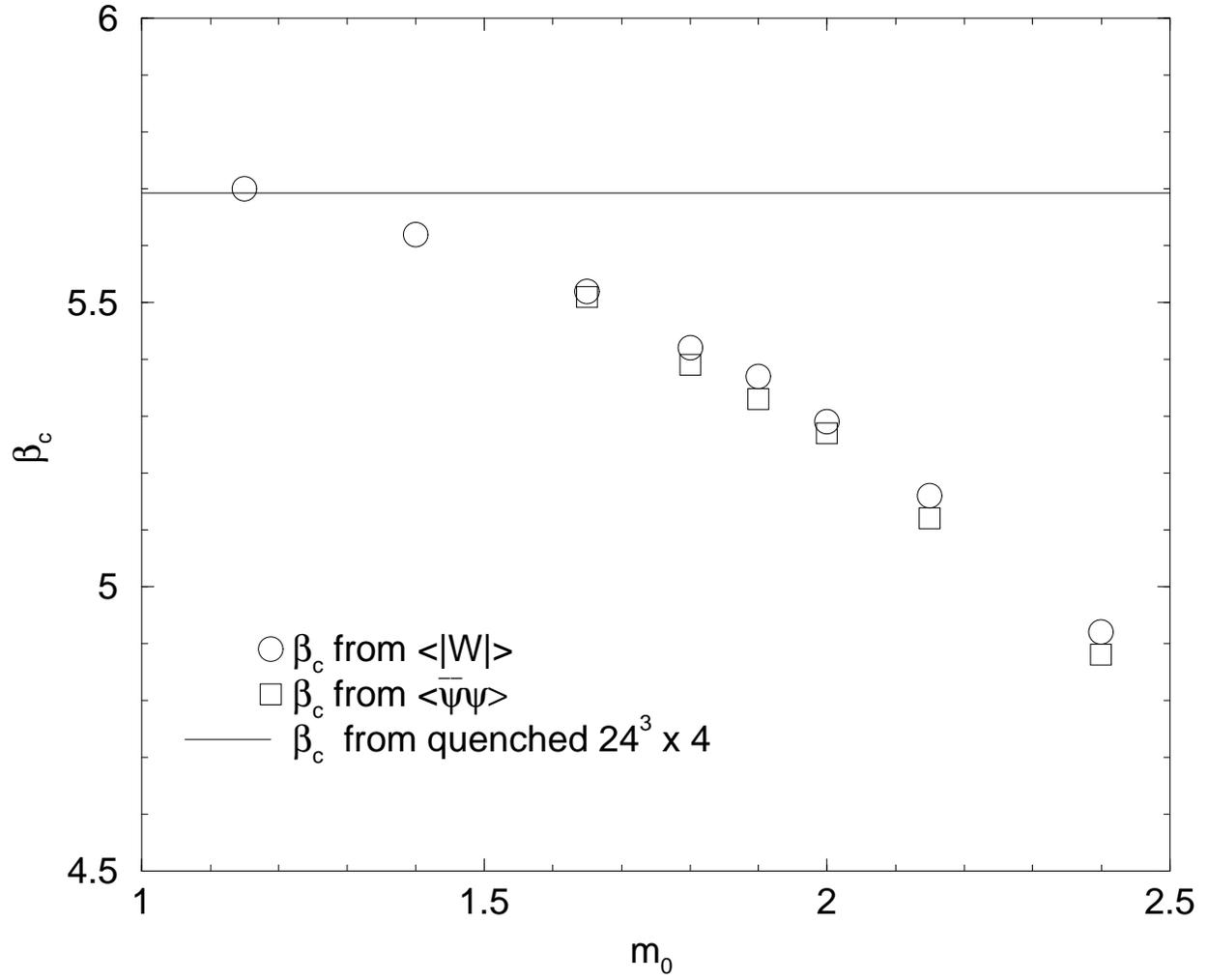}
\caption{ The critical value of $\beta$ as a function of $m_0$.  The
line is the value for a $24^3 \times 4$ lattice.}
\label{fig:betac_vs_m0}
\end{figure}

Figure \ref{fig:betac_vs_m0} gives estimates for $\beta_c$ determined
from the Wilson line and $\langle \bar{\psi} \psi \rangle$.  These are
in quite reasonable agreement, particularly given the phenomenological
character of their determination.  For $m_0 \sim 1.2$, $\beta_c$ is
close to the quenched value and moves smoothly to smaller values as
$m_0$ is increased.  For these larger values for $m_0$, the light quark
states appear and the maximum momentum for a state bound to the walls
should increase.  These light states make $\langle \bar{\psi} \psi
\rangle$ show crossover behavior and are required for our simulations
to be proper studies of two-flavor QCD.  At our largest value of 
$m_0$ (2.4), we may be approaching the transition from a two 
flavor theory to an eight flavor one (recall that the domain wall 
determinant is squared in our simulations, doubling the number of 
fermion flavors.)

\section{Determining the residual mass}
\label{sec:mres}

As mentioned in Section \ref{sec:analytical}, it can be expected that
for long-distance physical quantities, the effects of mixing between 
the chiral wall states will result in a residual mass contribution 
to the total quark mass.  This is just the statement that the 
dominant effect of the mixing, from the perspective of a low-energy
effective Lagrangian, is to introduce another source for
chiral symmetry breaking (beyond the input $m_f$), which takes the form
of the operator $m_{\rm res} \bar{\psi} \psi$ at low energies.  For a 
quantity like $m_\pi^2$, whose dependence on chiral symmetry
breaking can be expressed as a physical parameter times the total quark 
mass, the quark mass which enters should be $m_f + m_{\rm res}$.

However, for quantities whose sensitivity to chiral symmetry breaking
effects extends up to the cutoff scale, such an argument does not go 
through.  The chiral condensate, $\langle \bar{\psi} \psi \rangle$ 
is such a quantity.  For domain wall fermions with $L_s \rightarrow \infty$ 
(or staggered fermions), expanding in the input quark mass in the 
chirally broken phase gives
\begin{equation}
  \langle \bar{\psi} \psi \rangle = c_0 + c_1 m_f + O(m_f^2).
\end{equation}
The coefficient $c_1$ is ultraviolet divergent in the continuum and
therefore, on the lattice, gets large contributions from modes at
the cutoff scale.  For such an operator, the $L_s$ dependence is
not reliably represented by just making the replacement $m_f \rightarrow
m_f + m_{\rm res}$.

From this discussion, it is clear that although Figure
\ref{fig:b5_2_pbp_vs_mf_ls} shows that the large $L_s$ limit for
$\langle \bar{\psi} \psi \rangle$ at $m_f = 0.02$ has likely been
reached by $L_s \sim 40$, one cannot conclude that the value for
$m_{\rm res}$ has vanished.  To measure $m_{\rm res}$, it is natural to
look for effects in the pion mass, which is in turn governed by the
axial Ward-Takahashi identity.  This has been done in quenched
simulations 
Refs.~\cite{Blum:1997jf,Mawhinney:1998ut,Fleming:1998cc,Fleming:1999eq,Wu:1999cd,AliKhan:1999zn,Aoki:1999uv,Aoki:2000pc,Chen:1998xw}, at zero temperature, 
but here we are interested in determining $m_{\rm res}$ in the 
confined phase at finite temperature for small volumes for $N_f = 2$ QCD.

Our small volumes preclude taking large separations in two-point
functions to completely isolate the pion from other states.  Thus a
direct measurement of the pion mass or the overlap of the pion with any
particular source is not possible here.  Instead, we use the integrated
form for the flavor non-singlet axial Ward-Takahashi identity and try
to see the contributions of the pion.  In the zero quark mass limit on
infinite volumes, the pion contributions become poles.  Thus we can
look for the effects of these precursors of the pion poles, even when
they do not completely dominate the Ward-Takahashi identity.

Starting from the flavor non-singlet axial Ward-Takahashi identity
in \cite{Furman:1995ky} and summing over all lattice points gives
\begin{equation}
\label{eq:ward}
  \langle \bar{\psi} \psi \rangle = m_f \chi_\pi + \Delta J_5.
\end{equation}
Here $\psi$ is the four-dimensional fermion field defined by
Eq.\ \ref{projection} and the pseudoscalar susceptibility is (no sum on
$a$)
\begin{equation}
\chi_\pi \equiv \frac{2}{4 N_c} \sum_x \left\langle
  \bar{\psi}(x) \gamma_5 \frac{\lambda^a}{2} \psi(x)
  \, \bar{\psi}(0) \gamma_5 \frac{\lambda^a}{2} \psi(0)
\right\rangle,
\end{equation}
(The factor of $1/4N_c$ is needed to match our normalization for
$\langle \bar{\psi} \psi \rangle$.)
The additional contribution from chiral mixing due to finite $L_s$ is
\begin{equation}
\Delta J_5 \equiv \frac{2}{4 N_c} \sum_x \left\langle
  j_5^a\left(x,L_s/2\right)
  \, \bar{\psi}(0) \gamma_5 \frac{\lambda^a}{2} \psi(0)
\right\rangle,
\end{equation}
where
\begin{eqnarray}
  j_5^a(x, L_s/2 )
  & = &{1\over 4}\Psibar(x, L_s/2 ) (1 - \gamma_5 )\lambda^a \Psi(x, L_s/2 + 1 )
\\ \nonumber
  & - &{1\over 4}\Psibar(x, L_s/2 + 1 ) (1 + \gamma_5)\lambda^a\Psi(x, L_s/2 )
\end{eqnarray}
is a pseudoscalar density at the midpoint of the fifth dimension
which couples left- and right-handed degrees of freedom.

\begin{figure}
\epsfxsize=\hsize
\epsfbox{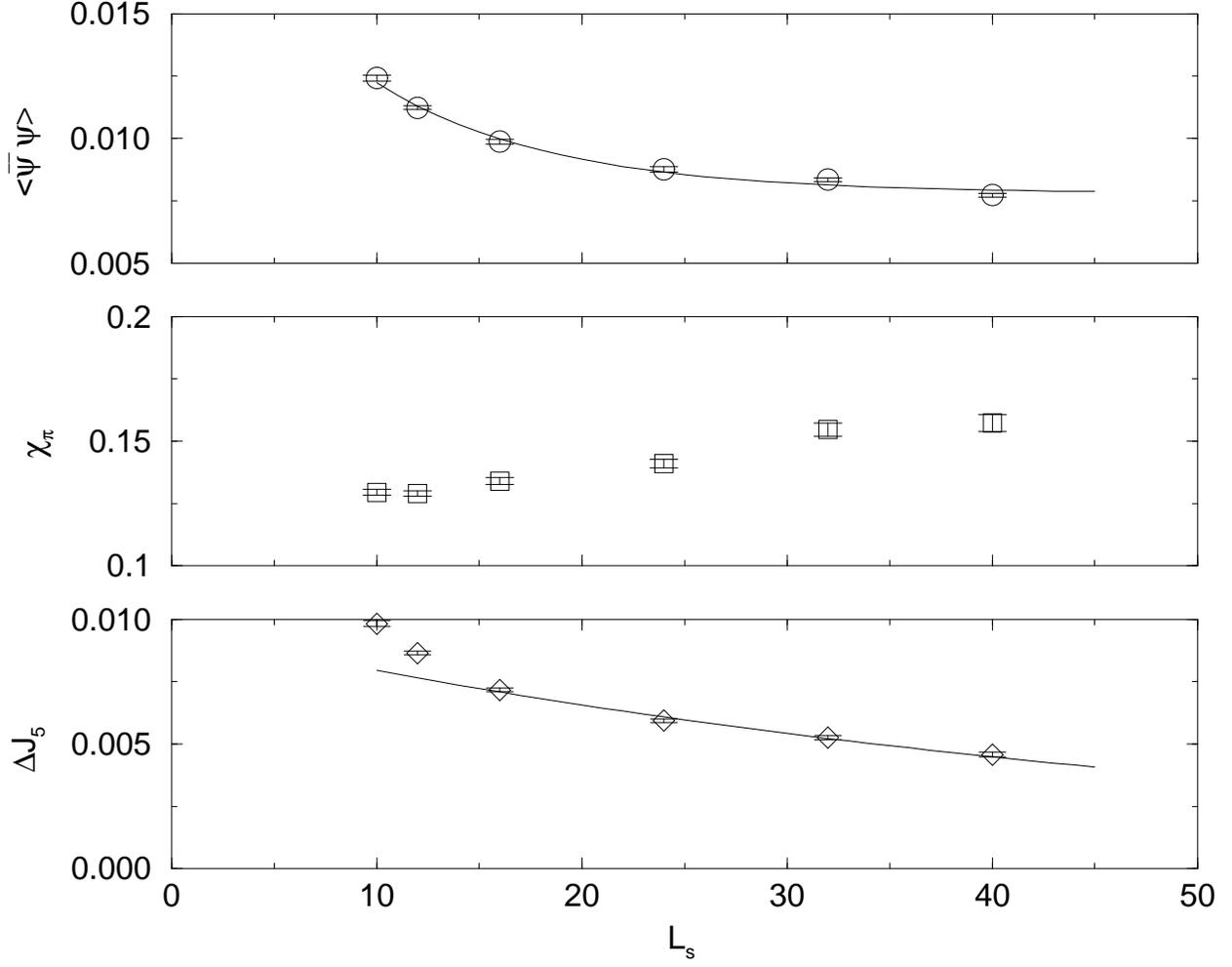}
\caption{The chiral condensate, $\langle \bar{\psi} \psi \rangle$,
the pion susceptibility, $\chi_\pi$ and $\Delta J_5$ versus $L_s$
for $\beta = 5.2$, $m_0 = 1.9$ and $m_f = 0.02$.}
\label{fig:pbp_spc_dj5_vs_ls}
\end{figure}

We have done extensive simulations for many values of $L_s$ with $\beta
= 5.2$, $m_0 = 1.9$ and $m_f = 0.02$ to study the consequences of the
Ward-Takahashi identity.  At the time of these simulations, we were not
measuring $\Delta J_5$ explicitly.  However, the other two terms in the
Ward-Takahashi identity were measured, allowing a determination of the
$\Delta J_5$ term.  Figure \ref{fig:pbp_spc_dj5_vs_ls} shows $\langle
\bar{\psi} \psi \rangle$, $\chi_\pi$ and $\Delta J_5$ for a variety of
values of $L_s$.  Fitting $\Delta J_5$ to an exponential form for $L_s
= 16$ to 40 gives the solid line in the figure and the result
\begin{equation}
  \Delta J_5 = 0.0096(2) \exp( -0.0191(9) L_s ) \,\,\,
  \chi^2/{\rm dof} = 6.4/2
\end{equation}
We see that our data is consistent with $\Delta J_5$ vanishing as
$L_s \rightarrow \infty$, although the decay constant is quite
small, $\approx 1/50$.

Pion poles should dominate the Ward-Takahashi identity when the
pions are light and the pions should become massless when
$m_f + m_{\rm res} = 0$.  (This is only strictly true in the infinite
volume limit.)  Thus we look for the pseudoscalar susceptibility
in large volumes for small total quark mass to behave as
\begin{equation}
\label{eq:chi_pi_ansatz}
\chi_\pi = a_{-1}/(m_f+m_{\rm res}) + a_0 + {\cal O}(m_f+m_{\rm res}).
\end{equation}
where the $a_{i}$ are independent of $L_s$ and $m_f$.  This gives
a pion pole (for large volumes) at $m_f = - m_{\rm res}$, while
$a_0$ gives the contribution to the susceptibility of modes whose
mass is non-zero when the quark mass vanishes. Like 
$\langle\psibar\psi\rangle$, $a_0$ receives contributions diverging
as $1/a^2$ and hence may be sensitive to unphysical 5-dimensional 
modes.  For this expression to be useful, we do not require the pole 
term to dominate the remaining terms, but it must make a large 
enough contribution to be visible.

The $\Delta J_5$ term in Eq.\ \ref{eq:ward} also has a pole
contribution coming from the propagation of the conventional light
pseudoscalar along the $s=0$ and $L_s-1$ boundaries from $0$ to $x$.
This light state has non-zero overlap with the midpoint pseudoscalar
density for finite $L_s$, but this overlap should be exponentially
suppressed.  Therefore we expect $\Delta J_5$ to also have a pole at
$m_f = - m_{\rm res}$, giving $\Delta J_5$ the same form as $\chi_\pi$,
namely
\begin{equation}
\Delta J_5 = b^\prime_{-1}/(m_f+m_{\rm res}) + b^\prime_0 +
  {\cal O}(m_f+m_{\rm res}).
\end{equation}

Considering the case where the pole terms dominate gives
\begin{equation}
 \langle \bar{\psi} \psi \rangle = \frac{a_{-1}m + b^\prime_{-1}}
   {m + m_{\rm res}}
\end{equation}
For $\langle \bar{\psi} \psi \rangle$ to be finite in this case
requires
\begin{equation}
  a_{-1}m + b^\prime_{-1} = a_{-1}( m + m_{\rm res} )
\end{equation}
so the most general form for $\Delta J_5$ is
\begin{equation}
\label{eq:dj5_ansatz}
\Delta J_5 = m_{\rm res}\chi_\pi + b_0 + {\cal O}(m_f+m_{\rm res}),
\end{equation}
Where $b_0=b_0^\prime - m_{\rm res} a_0$.   Using this then gives
\begin{equation}
  \label{eq:pbp_ansatz}
  \langle \bar{\psi} \psi \rangle = ( m_f + m_{\rm res} ) \chi_\pi + b_0
\end{equation}
up to terms linear in the quark mass.

Our procedure for extracting $m_{\rm res}$ from these small volumes
involves measuring values for $\chi_\pi$ and $\langle \bar{\psi} \psi
\rangle$ for a variety of valence quark masses for a simulation with a
fixed dynamical quark mass.  Since the Ward-Takahashi identity is a
consequence of the form of the domain wall fermion operator,
independent of the weight used to generate the gauge field ensemble in
which the fermionic observables are measured, it is satisfied
by observables measured with valence masses.  Of course, extrapolations
in valence quark mass can lead to problems due to the gauge field
ensemble including configurations with small fermion eigenvalues
that are not present when a dynamical extrapolation is done.  Here
we have a small dynamical quark mass present in the generation
of the gauge fields, so such effects are expected to be unimportant.

For a given $L_s$, we simultaneously fit $\chi_\pi$ and $ \langle
\bar{\psi} \psi \rangle$ to the forms in Eqs.\ \ref{eq:chi_pi_ansatz}
and \ref{eq:pbp_ansatz}.  These are four parameter fits for
$a_0, a_{-1}, b_0$ and $m_{\rm res}$ and the resulting value for
$m_{\rm res}$ we refer to as $m^{(\rm GMOR)}_{\rm res}$.  (All
measurements of the residual mass from low energy physics should
agree.  We use this notation to detail the explicit technique
we have used for this determination.)  We have used quark
masses of 0.02, 0.06, 0.10 and 0.14 in our fits.  These fits do not
include possible correlations between the quantities computed for
different values of $m_f$ because the correlation matrix itself is 
poorly determined.

\begin{figure}
\epsfxsize=\hsize
\epsfbox{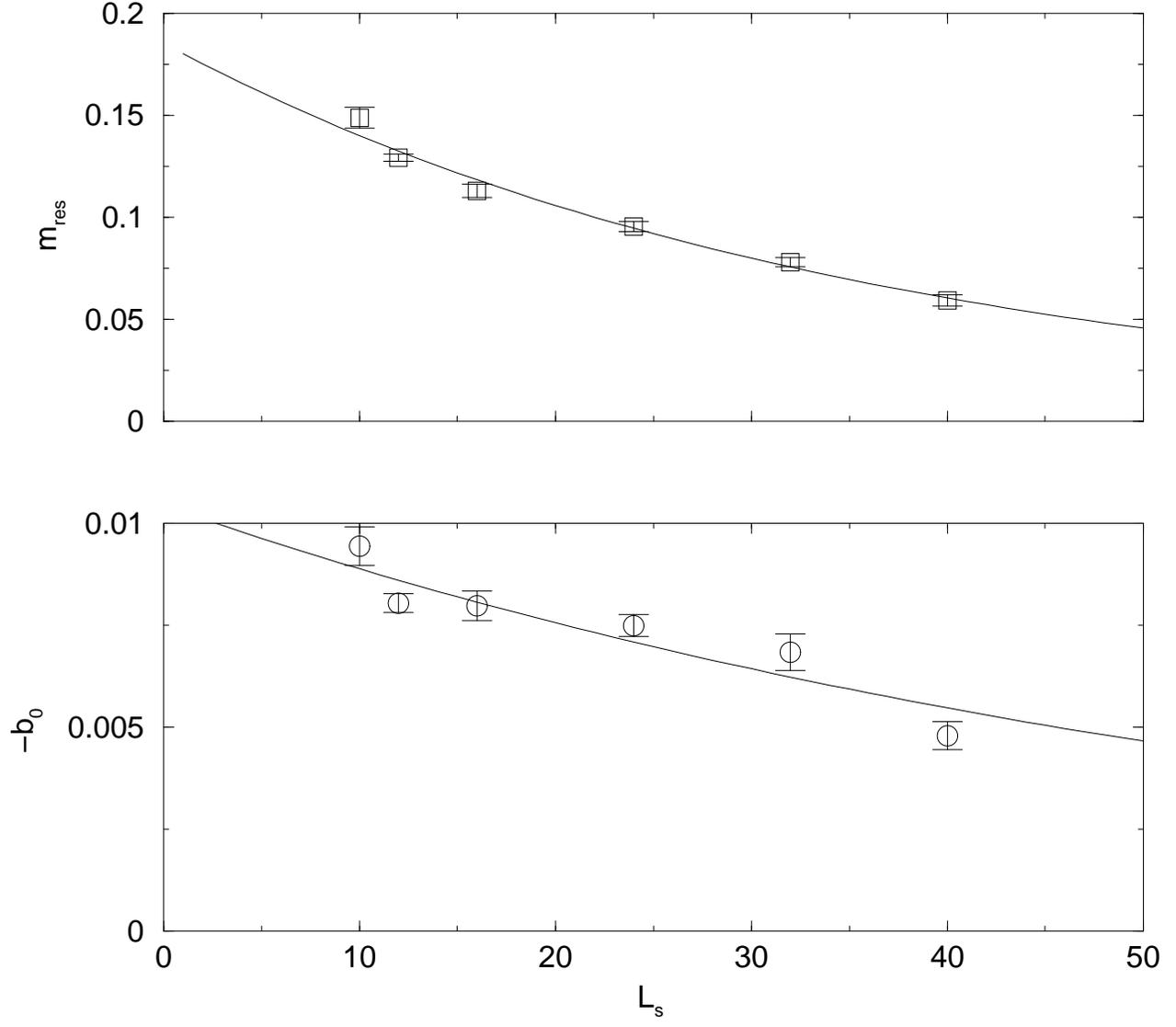}
\caption{ $m^{(\rm GMOR)}_{\rm res}$ and $-b_0$ versus $L_s$.  The
curves are fits to the form $c_0 + c_1 \exp( -\alpha L_s)$. } 
\label{fig:mres_b0_vs_ls}
\end{figure}

The results are given in Table \ref{tab:mres_vs_ls}, where the errors
are all from application of the jack knife method.  Notice that $b_0$ is negative for all
values of $L_s$, meaning that the non-pole contributions to
$\Delta J_5$ are smaller than $m_{\rm res} a_0$.  We have then fit
these values of $m^{(\rm GMOR)}_{\rm res}$ and $-b_0$ to the form
$c_0 + c_1 \exp( -\alpha L_s)$ and found
\begin{eqnarray}
 -b_0 & = & 0.0104(4) \exp( -0.016(2) L_s )
   \,\,\, \chi^2/{\rm dof} = 0.34(19)\\
  m^{(\rm GMOR)}_{\rm res} & = & 0.185(6) \exp( -0.0280(15) L_s )
   \,\,\,\,\, \chi^2/{\rm dof} = 0.28(25)\\
\end{eqnarray}
Figure \ref{fig:mres_b0_vs_ls} shows these values and the fits.

We can see that both $m^{(\rm GMOR)}_{\rm res}$ and $b_0$ are falling
exponentially, but with a very small decay constant $\approx 1/50$.
This is in sharp contrast to the decay constant for $\langle \bar{\psi}
\psi \rangle$ which is $\approx 1/10$.  This is further evidence for
the distinction between the residual mass that enters in low-energy
observables and the residual mixing which effects observables
dependent on degrees of freedom at the cutoff scale.

Since our determination of the residual mass has been done for small
volumes, one can worry about the finite volume effects.  We have done a
similar extraction of the residual mass and compared it with
determinations of the residual mass from extrapolations of $m_\pi^2$
for much larger volumes and find reasonable agreement 
\cite{Fleming:1999eq}.  We are continuing to study various determinations 
of the residual mass.

\section{Conclusions}
\label{sec:conclusion}

In this work the properties of domain wall fermions relevant to
numerical simulations of full $N_f = 2$ QCD at finite temperature were
investigated on relatively small lattices of size $8^3 \times 4$.
Conventional numerical algorithms (the Hybrid Monte Carlo and the
conjugate gradient) worked without any difficulty beyond the additional
computational load of the fifth dimension.  Evidence for both confined
and deconfined phases was found and the $L_s$ and $m_0$ dependence of
each phase was investigated.

The domain wall fermion action is expected to preserve the full
chiral symmetries of QCD for large $L_s$.  For the stronger
couplings used for the confined phase simulations, the chiral
condensate approached its asymptotic value for $L_s \approx 32-40$.
However, our determination of the residual mass effects present
in low energy observables show a residual mass of $\approx 0.06$
for $L_s = 40$.  For the weaker couplings needed to study the
deconfined, chirally restored phase, the residual mass effects
are expected to be much smaller for the same $L_s$, although we
have not yet measured the residual mass in this region.

In particular, it was found that for the two flavor theory there is a
phase where the $SU(2) \times SU(2)$ chiral symmetry is broken
spontaneously to a full $SU(2)$ flavor symmetry and a phase where the
full $SU(2) \times SU(2)$ chiral symmetry is intact.  For the values of
$L_s$ we used, the dependence of observables on the coupling in the
transition region is likely quite influenced by the change in the
residual mass with the coupling.  To suppress this effect
will require larger values for $L_s$, thermodynamics studies at larger
$N_t$ (and hence weaker coupling) or improved variants of domain wall
fermions.

Our simulations show that domain wall fermions have passed one vital
test for numerical work, light chiral modes exist at quite strong
coupling.  A second important result, which was expected from work with
dynamical fermions in the Schwinger model \cite{Vranas:1997da}, is that
domain wall fermions do not present any problems to conventional
dynamical fermion numerical algorithms.  Given these results, we are
pursuing simulations of the phase transition on larger lattices to
achieve more physically meaningful results.  The slow falloff of the
residual mass with $L_s$ can be overcome with more computing power or,
hopefully, improvements to the formulation.  At present, this is all
that stands in the way of simulating the $N_f=2$ QCD phase transition
with three degenerate light pions at finite lattice spacing.


\section*{Acknowledgments}

The numerical calculations were done on the 400 Gflop QCDSP computer
\cite{Chen:1998cg} at Columbia University. This research was supported
in part by the DOE under grant \# DE-FG02-92ER40699
and for P. Vranas in part by NSF under grant \# NSF-PHY96-05199.


\appendix
\section{Gamma matrices}
\label{appendix_gamma}
The Dirac gamma matrices used in this work are:
\begin{eqnarray}
&&
\gamma_1 = \left( \begin{array}{cccc}  ~~~0 &  ~~~0 &  ~~~0 &  ~~~i \\ 
                                       ~~~0 &  ~~~0 &  ~~~i &  ~~~0 \\
                                       ~~~0 &    -i &  ~~~0 &  ~~~0 \\
                                         -i &  ~~~0 &  ~~~0 &  ~~~0 \end{array} \right), 
\ \  
\gamma_2 = \left( \begin{array}{cccc}  ~~~0 &  ~~~0 &  ~~~0 &    -1 \\ 
                                       ~~~0 &  ~~~0 &  ~~~1 &  ~~~0 \\
                                       ~~~0 &  ~~~1 &  ~~~0 &  ~~~0 \\
                                         -1 &  ~~~0 &  ~~~0 &  ~~~0 \end{array} \right), 
\nonumber \\  &&
\gamma_3 = \left( \begin{array}{cccc}  ~~~0 &  ~~~0 &  ~~~i &  ~~~0 \\ 
                                       ~~~0 &  ~~~0 &  ~~~0 &    -i \\
                                         -i &  ~~~0 &  ~~~0 &  ~~~0 \\
                                       ~~~0 &  ~~~i &  ~~~0 &  ~~~0 \end{array} \right), 
\ \  
\gamma_4 = \left( \begin{array}{cccc}  ~~~0 &  ~~~0 &  ~~~1 &  ~~~0 \\ 
                                       ~~~0 &  ~~~0 &  ~~~0 &  ~~~1 \\
                                       ~~~1 &  ~~~0 &  ~~~0 &  ~~~0 \\
                                       ~~~0 &  ~~~1 &  ~~~0 &  ~~~0 \end{array} \right), 
\nonumber \\  &&
\gamma_5 = \left( \begin{array}{cccc}  ~~~1 &  ~~~0 &  ~~~0 &  ~~~0 \\ 
                                       ~~~0 &  ~~~1 &  ~~~0 &  ~~~0 \\
                                       ~~~0 &  ~~~0 &    -1 &  ~~~0 \\
                                       ~~~0 &  ~~~0 &  ~~~0 &    -1 \end{array} \right) .
\end{eqnarray}
%


\section{Evolution Algorithm}
\label{appendix_md}

As described in Section~\ref{sec:hmc}, we use the Hybrid Monte Carlo 
`$\Phi$' algorithm of Gottlieb {\it et al.}\cite{Gottlieb:1987mq} 
extended to include the Pauli-Villars regulator fields.  Further, we use a 
preconditioned variant of the Dirac operator specified in Eq.~\ref{eq:D_F}\cite{DeGrand:1988vx}.  In this Appendix we describe
the resulting algorithm we use to evolve the gauge fields including the
effects of the two flavors of domain wall quarks and the Pauli-Villars 
regulator fields.

Following this approach, we generate a Markov chain of gauge fields 
$U_\mu(x)$, pseudo-fermion fields $\Phi_{F}$, Pauli-Villars
fields $\Phi_{PV}$ and conjugate momenta $H_\mu(x)$ according to 
the distribution:
\begin{equation}
Z = \int[dU][dH][d\Phi_{F}^\dagger][d\Phi_{F}]
        [d\Phi_{PV}^\dagger][d\Phi_{PV}] e^{-{\cal H}}
\end{equation}
where
\begin{equation}
{\cal H} = S_G
  \;+{1\over 2}\sum_{x,\mu}H_\mu(x)^2 
  \;+\Phi_{F}^\dagger[\widetilde{D}_F^\dagger\widetilde{D}_F]^{-1}\Phi_{F}
  \;+\Phi_{PV}^\dagger [\widetilde{D}_F^\dagger\widetilde{D}_F]_{m_f=1}\Phi_{PV}.
\label{eq:H_6}
\end{equation}
Here, the fields $\Phi_F$ and $\Phi_{PV}$ as well as the preconditioned operator 
$\widetilde{D}_F$ are defined only on odd sites with
\begin{equation}
\widetilde{D}_F = (5-m_0)^2-(D_F)_{oe}(D_F)_{eo}
\label{eq:preconditioning}
\end{equation}
where $(D_F)_{oe}$ and $(D_F)_{eo}$ represent the DWF operator of 
Eq.~\ref{eq:D_F} evaluated between odd and even or even and odd sites 
respectively.  Note, even and odd are defined in a five-dimensional 
sense, {\it e.g.} for an even site the sum of all five coordinates is an 
even number.  Eq.~\ref{eq:preconditioning} employs the usual 
preconditioning scheme for Wilson fermions\cite{DeGrand:1988vx} 
implemented in 5 dimensions. Similar considerations justify the form
used for the Pauli-Villars action.  Since $\det\widetilde{D}_F
=\det\{(5-m_0)D_F\}$, we have rescaled both the fields
$\Phi_F$ and $\Phi_{PV}$ to introduce the extra factor of $(5-m_0)$ into 
Eq.~\ref{eq:preconditioning} in order to simplify the subsequent algebra.

To begin a new HMC trajectory, we start with the values of the gauge 
fields $U_\mu(x)$ produced by the previous trajectory.  We then choose 
Gaussian distributed fields $\eta(x,s)_F$, $\eta(x,s)_{PV}$ and 
$H_\mu(x)$ from which we construct the fields 
$\Phi_F = \widetilde{D}_F \eta_F$ and 
$\Phi_{PV} = (\widetilde{D}_F^{-1}|_{m_f=1})\eta_{PV}$.  
Here we have introduced new field variables $H_\mu(x)$, conjugate to 
the link matrices, which are elements of the algebra of $SU(3)$, and 
hence traceless and hermitian.

Next, we carry out the molecular dynamics time evolution of the fields 
$H_\mu(x)$ and $U_\mu(x)$ according to equations of motion which are 
phase space volume preserving and conserve the fictitious 6-dimensional 
``energy'' $\cal H$ of Eq.~\ref{eq:H_6}.  The first of these 
Hamilton-like equations determines the relation between $U_\mu(x)$ 
and the conjugate variable $H_\mu(x)$:
\begin{equation}
\frac{{\rm d}U_\mu(x)}{{\rm d}\tau} = i H_\mu(x) U_\mu(x).
\end{equation}
The second equation can be derived from the requirement that ${\cal H}$
is $\tau$-independent.  First, following Gottlieb 
{\it et al.}\cite{Gottlieb:1987mq} one writes:
\begin{equation}
\frac{{\rm d}{\cal H}}{{\rm d}\tau}
= \sum_{x,\mu}{\rm Tr}\left[ i H_\mu(x) F_\mu(x) 
     + \frac{{\rm d}H_\mu(x)}{{\rm d}\tau} H_\mu(x) \right].
\label{eq:F_def}
\end{equation}
Then the constancy of $\cal H$ is insured if for the second 
equation of motion we impose:
\begin{equation}
i \frac{{\rm d}H_\mu(x)}{{\rm d}\tau}
= \left[ F_\mu(x) \right]_{TA}.
\end{equation}
The subscript $TA$ indicates the traceless anti-hermitian 
part of the matrix, a restriction required by the traceless, hermitian
character of the variables $H_\mu(x)$.  (The definition of $F_\mu(x)$ 
implied by Eq.~\ref{eq:F_def} makes $F$ anti-hermitian and it is
only the traceless part of $F$ that enters that equation.)

Finally we will determine the specific form for the force term 
$F_\mu(x)$.  This can be done by using the general formula
\begin{eqnarray}
\label{eq:d_by_d_tau}
{{\rm d} \over {\rm d} \tau} \langle\psi\,^\prime|D_F^{(\dagger)}|\psi\rangle &=&
   {i\over2}\sum_{x,s} 
\Big\{\psi\,^\prime(x,s)^\dagger H_\mu(x)U_\mu(x)(1\mp\gamma^\mu)\psi(x+\mu,s) 
\label{eq:eval_derivative} \\ 
&&\quad
-\psi\,^\prime(x+\mu,s)^\dagger U_\mu(x)^\dagger H_\mu(x)(1\pm\gamma^\mu)\psi(x,s)\Big\}
\nonumber
\end{eqnarray}
which follows immediately from Eq's.~\ref{eq:D_parallel} and \ref{eq:F_def}
where the lower choice of signs corresponds to the case of $D_F^\dagger$.
Now we re-express the derivative:
\begin{equation}
{{\rm d} \over {\rm d} \tau} 
   \Phi_F^\dagger [\widetilde{D}_F^\dagger \widetilde{D}_F]^{-1}\Phi_F =
   -\chi_F^\dagger\left[{{\rm d} \over {\rm d} \tau}\widetilde{D}_F^\dagger          \widetilde{D}_F\right]\chi_F,
\label{eq:dif_denom}
\end{equation}
where we construct $\Phi=\widetilde{D}_F \eta_F$ from the Gaussian
source $\eta_F$ and then obtain $\chi_F$ by solving $\widetilde{D}_F^\dagger\widetilde{D}_F \chi_F = \Phi_F$.
Now we must evaluate
\begin{eqnarray}
\chi_F^\dagger\left[{{\rm d} \over {\rm d} \tau}\widetilde{D}_F^\dagger          \widetilde{D}_F\right]\chi_F 
&=& {{\rm d} \over {\rm d} \tau} 
\langle\chi_F|[(5-m_0)^2-(D_F^\dagger)_{oe}(D_F^\dagger)_{eo}] 
\label{eq:expand_D_tilde} \\
&&\quad\cdot[(5-m_0)^2-(D_F)_{oe}(D_F)_{eo}] |\chi_F\rangle,
\nonumber
\end{eqnarray}
We will obtain eight terms by letting the derivative act on each of the 
four $D_F$ operators.  Four of those terms will involve $U_\mu(x)$ and 
four $U_\mu(x)^\dagger$, with the final four terms being the hermitian 
conjugates of the first four.  Combining Eq.'s~\ref{eq:F_def}, 
\ref{eq:eval_derivative}, \ref{eq:dif_denom} and \ref{eq:expand_D_tilde}, 
we find:
\begin{eqnarray}
{\rm tr}\{ H_\mu(x) F_\mu(x)\} &=& 
 {1\over 2}\sum_s\Big\{\chi_F(x,s)H_\mu(x)U_\mu(x)(1+\gamma^\mu)\langle(x+\mu,s)
                  |(D_F^\dagger)_{eo}\widetilde{D}_F|\chi_F\rangle \nonumber\\
  &+&\quad\langle \chi_F|(D_F^\dagger)_{oe}|x,s\rangle H_\mu(x)U_\mu(x)
     (1+\gamma^\mu)\langle x+\mu,s|\widetilde{D}_F|\chi_F\rangle  \nonumber\\
  &+&\quad\langle \chi_F|\widetilde{D}_F^\dagger|x,s\rangle H_\mu(x)U_\mu(x)
     (1-\gamma^\mu)\langle x+\mu,s|(D_F)_{eo}|\chi_F\rangle           \nonumber\\
  &+&\quad\langle \chi_F|\widetilde{D}_F^\dagger (D_F)_{oe}|x,s\rangle 
     H_\mu(x)U_\mu(x)(1-\gamma^\mu)\chi(x+\mu,s) - {\rm h.c.} \Big\}.
\label{eq:force}
\end{eqnarray}
This expression can be written in a very simple form if we define two new
spinor quantities:
\begin{eqnarray}
w(x,s)&=&\left\{ \begin{array}{ll} 
 -\langle x,s|(D_F^\dagger)_{eo}\widetilde{D}_F |\chi_F\rangle 
                                                  &\quad\mbox{$(x,s)$ even} \\
 -\langle x,s|\widetilde{D}_F|\chi_F\rangle                  
                                                  &\quad\mbox{$(x,s)$ odd}
                 \end{array}  \right. \\
v(x,s)&=&\left\{ \begin{array}{ll}
    \langle x,s|(D_F)_{eo}|\chi_F\rangle  &\quad\quad\mbox{$(x,s)$ even}\\  
    \chi_F(x,s)                           &\quad\quad\mbox{$(x,s)$ odd}
                  \end{array} \right.
\end{eqnarray}
Using these quantities in Eq.~\ref{eq:force} and factoring out the 
generator $H_\mu(x)$ gives:
\begin{eqnarray}
F_{[F]\mu}(x) &=& -{1\over 2} U_\mu(x) \sum_s {\rm tr}_{\rm spin}\;\Big[
  (1-\gamma_\mu) v(x+\hat\mu,s) w^\dagger(x,s) \nonumber\\
  &&\quad + (1+\gamma_\mu) w(x+\hat\mu,s) v^\dagger(x,s) \Big] -{\rm h.c.}
\end{eqnarray}
where we have added now the subscript $[F]$ to distinguish this fermion 
force from that produced by the Pauli-Villars fields described below.  
Since there are no gauge fields in the extra direction, it is not 
surprising that this looks very similar to the Wilson fermion force 
with an additional sum over the s-direction.

The force term produced by the Pauli-Villars fields is closely related
to that derived above.  We need only replace the field $\chi_F$ with
$\Phi_{PV}$, set $m_f=1$ and change the sign of the resulting force:
\begin{equation}
F_{[PV]\mu}(x) = - \left. F_{[F]\mu}(x) \right|_{m_f=1, \chi_F=\Phi_{\rm PV}}.
\end{equation}


\begin{figure}
\epsfxsize=\hsize
\epsfbox{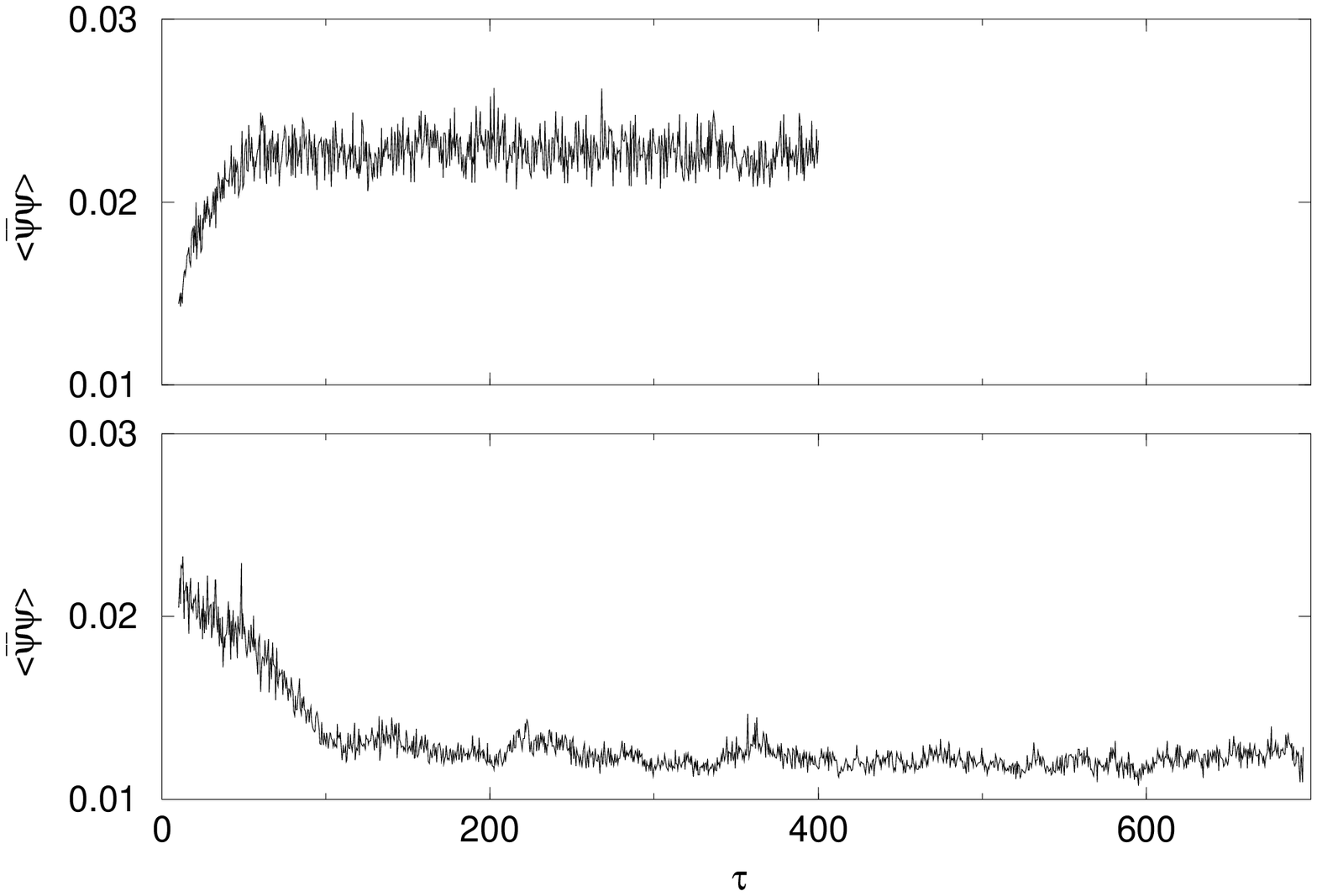}
\caption{ The simulation ``time history" of $\langle \bar{\psi} \psi
\rangle$ for $\beta = 5.0$ (upper graph) and $\beta = 5.4$ lower graph
for $m_f = 0.1$, $m_0 = 1.90$ and $L_s = 8$.}
\label{fig:b5_0_b5_4_ls8_M1_90_evol}
\end{figure}

\begin{table}
  \centering
  \begin{tabular}{ccccccccc}
    $m_f$ & $L_s$ & traj.~len. & \# traj. & acc. &
    $\left\langle e^{-\Delta H}\right\rangle$ & $\plaq$ & $\w3m$ & $\pbp$ \\
    \hline\hline
    0.02 &  8 & $\frac{1}{64}\times 32$   & 400-800  & 0.89 & 0.98(1)  &
      0.456(2)  & 0.061(2) & 0.0133(2)  \\
         & 10 & $\frac{1}{64}\times 32$   & 200-2000 & 0.86 & 0.995(8) &
      0.4460(9) & 0.048(2) & 0.0124(1)  \\
         & 12 & $\frac{1}{64}\times 32$   & 200-2000 & 0.84 & 1.01(1)  &
      0.4428(6) & 0.048(2) & 0.01123(7) \\
         & 16 & $\frac{1}{64}\times 32$   & 550-2000 & 0.75 & 0.98(2)  &
      0.4388(9) & 0.049(2) & 0.00987(9) \\
         & 24 & $\frac{1}{100}\times 50$  & 350-2000 & 0.73 & 0.95(2)  &
      0.4359(7) & 0.047(3) & 0.0088(1)  \\
         & 32 & $\frac{1}{100}\times 50$  & 300-2000 & 0.72 & 1.03(2)  &
      0.4317(7) & 0.045(2) & 0.00835(7) \\
         & 40 & $\frac{1}{128}\times 64$  & 300-1350 & 0.73 & 1.02(3)  &
      0.4342(6) & 0.044(2) & 0.00772(8) \\
    \hline
    0.06 &  8 & $\frac{1}{50}\times 25$   & 200-950  & 0.83 & 0.98(1)  &
      0.450(1)  & 0.046(3) & 0.0176(2)  \\
         & 16 & $\frac{1}{64}\times 32$   & 200-820  & 0.84 & 0.99(2)  &
      0.4361(8) & 0.045(3) & 0.0135(1)  \\
    \hline
    0.1  &  8 & $\frac{1}{40}\times 20$   & 300-800  & 0.57 & 0.89(5)  &
      0.4437(8) & 0.040(3) & 0.02109(7) \\
         & 10 & $\frac{1}{50}\times 25$   & 200-800  & 0.83 & 1.00(3)  &
      0.4405(6) & 0.036(2) & 0.01927(9) \\
         & 12 & $\frac{1}{40}\times 20$   & 400-800  & 0.43 & 1.2(3)   &
      0.437(1)  & 0.032(2) & 0.01838(4) \\
         & 16 & $\frac{1}{64}\times 32$   & 200-800  & 0.80 & 0.98(2)  &
      0.435(2)  & 0.035(2) & 0.01709(9) \\
         & 24 & $\frac{1}{64}\times 32$   & 200-800  & 0.72 & 0.94(2)  &
      0.433(1)  & 0.033(2) & 0.01596(7) \\
         & 32 & $\frac{1}{100}\times 50$  & 200-800  & 0.82 & 0.99(2)  &
      0.4305(5) & 0.037(2) & 0.01547(7) \\
         & 40 & $\frac{1}{100}\times 50$  & 200-800  & 0.78 & 1.00(5)  &
      0.432(1)  & 0.035(2) & 0.01524(5) \\
    \hline
    0.14 &  8 & $\frac{1}{40}\times 20$   & 200-860  & 0.63 & 1.03(6)  &
      0.4433(7) & 0.033(5) & 0.0241(1)  \\
         & 16 & $\frac{1}{64}\times 32$   & 200-800  & 0.85 & 1.03(2)  &
      0.433(1)  & 0.030(1) & 0.02017(6) \\
    \hline
    0.18 &  8 & $\frac{1}{40}\times 20$   & 200-1200 & 0.70 & 1.02(3)  &
      0.4410(7) & 0.030(1) & 0.02686(7) \\
         & 16 & $\frac{1}{64}\times 32$   & 200-800  & 0.84 & 0.98(2)  &
      0.432(1)  & 0.033(1) & 0.02309(5) \\
  \end{tabular}
  \caption{Data for an $8^3\times4$ lattice with $\beta=5.2$, $m_0=1.9$}
  \label{tab:8nt4_b5.2_h1.9}
\end{table}

\vfill
\null
\eject
\begin{table}
  \centering
  \begin{tabular}{ccccccccc}
    $m_f$ & $L_s$ & traj.~len. & \# traj. & acc. &
    $\left\langle e^{-\Delta H}\right\rangle$ & $\plaq$ & $\w3m$ & $\pbp$ \\
    \hline\hline
   0.02 &  8 & $\frac{1}{64}\times 32$  & 200-800  & 0.91 & 1.005(7) &
      0.5376(7) & 0.226(4) & 0.00415(6) \\
         & 10 & $\frac{1}{64}\times 32$  & 200-1000 & 0.91 & 0.992(8) &
      0.5328(6) & 0.207(4) & 0.00319(5) \\
         & 12 & $\frac{1}{64}\times 32$  & 200-800  & 0.95 & 1.009(9) &
      0.5300(4) & 0.202(5) & 0.00270(3) \\
         & 16 & $\frac{1}{64}\times 32$  & 200-800  & 0.90 & 1.02(1)  &
      0.5266(8) & 0.199(4) & 0.00237(6) \\
         & 24 & $\frac{1}{64}\times 32$  & 400-1200 & 0.86 & 0.98(2)  &
      0.5257(7) & 0.187(3) & 0.00216(6) \\
         & 32 & $\frac{1}{100}\times 50$ & 400-800  & 0.94 & 1.00(2)  &
      0.524(2)  & 0.180(5) & 0.00209(5) \\
    \hline
    0.06 &  8 & $\frac{1}{50}\times 25$  & 200-1000 & 0.86 & 0.99(3)  &
      0.536(1)  & 0.217(3) & 0.0080(1)  \\
         & 10 & $\frac{1}{64}\times 32$  & 200-1000 & 0.92 & 0.994(7) &
      0.5313(6) & 0.203(4) & 0.00704(5) \\
         & 12 & $\frac{1}{64}\times 32$  & 200-1000 & 0.89 & 1.013(8) &
      0.5286(8) & 0.195(4) & 0.00666(5) \\
         & 16 & $\frac{1}{64}\times 32$  & 400-800  & 0.76 & 1.02(4)  &
      0.525(2)  & 0.192(4) & 0.00637(7) \\
         & 24 & $\frac{1}{64}\times 32$  & 300-1000 & 0.84 & 1.00(1)  &
      0.521(2)  & 0.174(6) & 0.00617(9) \\
         & 32 & $\frac{1}{64}\times 32$  & 500-1000 & 0.80 & 1.00(2)  &
      0.525(2)  & 0.189(3) & 0.00592(4) \\
    \hline
    0.1  &  8 & $\frac{1}{50}\times 25$  & 300-800  & 0.83 & 0.98(2)  &
      0.5336(6) & 0.211(4) & 0.01174(4) \\
         & 10 & $\frac{1}{50}\times 25$  & 300-990  & 0.88 & 0.99(1)  &
      0.5310(9) & 0.200(2) & 0.01075(5) \\
         & 12 & $\frac{1}{50}\times 25$  & 600-1200 & 0.74 & 1.01(4)  &
      0.528(1)  & 0.197(4) & 0.01838(4) \\
         & 16 & $\frac{1}{64}\times 32$  & 400-800  & 0.79 & 1.01(3)  &
      0.523(1)  & 0.170(5) & 0.0103(1)  \\
         & 24 & $\frac{1}{64}\times 32$  & 400-2000 & 0.86 & 0.991(8) &
      0.512(1)  & 0.170(8) & 0.0102(1)  \\
         & 32 & $\frac{1}{64}\times 32$  & 300-1000 & 0.81 & 0.98(2)  &
      0.519(1)  & 0.159(5) & 0.01011(9) \\
    \hline
    0.14 &  8 & $\frac{1}{50}\times 25$  & 200-800  & 0.83 & 1.01(1)  &
      0.533(1)  & 0.210(3) & 0.01531(9) \\
         & 16 & $\frac{1}{64}\times 32$  & 600-1200 & 0.76 & 0.98(2)  &
      0.520(1)  & 0.159(9) & 0.0143(1)  \\
    \hline
    0.18 &  8 & $\frac{1}{50}\times 25$  & 400-800  & 0.81 & 1.03(2)  &
      0.5314(6) & 0.202(4) & 0.01884(5) \\
         & 16 & $\frac{1}{64}\times 32$  & 600-1200 & 0.78 & 0.94(2)  &
      0.515(1)  & 0.141(8) & 0.0182(2)  \\
  \end{tabular}
  \caption{Data for an $8^3\times4$ lattice with $\beta=5.45$, $m_0=1.9$}
  \label{tab:8nt4_b5.45_h1.9}
\end{table}

\vfill
\null
\eject

\begin{table}
  \centering
  HMC traj.~len: $\frac{1}{50}\times 25$, \qquad CG stop cond: $10^{-6}$
  \begin{tabular}{cccccccc}
    $\beta$ & start & \# traj & acc. &
    $\left\langle e^{-\Delta H}\right\rangle$ & $\plaq$ & $\w3m$ & $\pbp$ \\
    \hline\hline
    5.45 & O & 100-800 & 0.87 & 0.99(1)  & 0.470(1)  & 0.0168(6) &
      0.00276(1)  \\
    5.55 & O & 200-800 & 0.87 & 0.98(1)  & 0.4933(6) & 0.023(1)  &
      0.002916(6) \\
    5.65 & O & 300-800 & 0.87 & 1.00(2)  & 0.5218(9) & 0.054(6)  &
      0.00305(2)  \\
    5.75 & D & 300-800 & 0.86 & 0.986(9) & 0.5571(7) & 0.196(7)  &
      0.002875(7) \\
    5.85 & D & 300-800 & 0.85 & 1.00(1)  & 0.5719(7) & 0.234(3)  &
      0.002881(3) \\
    5.95 & D & 200-800 & 0.87 & 0.99(1)  & 0.5857(5) & 0.262(2)  &
      0.002898(3)
  \end{tabular}
  \caption{Data for an $8^3\times 4$ lattice with $m_0=1.15$, $L_s=12$, and $m_f=0.1$.}
  \label{tab:8nt4_h1.15_beta_crit}
\end{table}

\begin{table}
  \centering
  HMC traj.~len: $\frac{1}{50}\times 25$, \qquad CG stop cond: $10^{-6}$ \\
  \begin{tabular}{cccccccc}
    $\beta$ & start & \# traj & acc. &
    $\left\langle e^{-\Delta H}\right\rangle$ & $\plaq$ & $\w3m$ & $\pbp$ \\
    \hline\hline
    5.35 & O & 100-800 & 0.87 & 1.01(1)  & 0.4435(7) & 0.0189(8) &
      0.00497(1)  \\
    5.45 & O & 100-800 & 0.86 & 0.99(1)  & 0.4630(7) & 0.0215(8) &
      0.00522(1)  \\
    5.55 & O & 300-800 & 0.86 & 1.00(1)  & 0.487(1)  & 0.032(3)  &
      0.00539(3)  \\
    5.65 & D & 400-800 & 0.86 & 1.00(2)  & 0.540(2)  & 0.180(6)  &
      0.00457(4)  \\
    5.75 & D & 300-800 & 0.85 & 0.99(1)  & 0.5598(8) & 0.224(3)  &
      0.00445(1)  \\
    5.85 & D & 200-800 & 0.89 & 1.011(8) & 0.5744(3) & 0.254(3)  &
      0.004409(5)
  \end{tabular}
  \caption{Data for an $8^3\times 4$ lattice with $m_0=1.4$, $L_s=12$, and $m_f=0.1$.}
  \label{tab:8nt4_h1.4_beta_crit}
\end{table}

\begin{table}
  \centering
  HMC traj.~len: $\frac{1}{50}\times 25$, \qquad CG stop cond: $10^{-6}$
  \begin{tabular}{cccccccc}
    $\beta$ & start & \# traj & acc. &
    $\left\langle e^{-\Delta H}\right\rangle$ & $\plaq$ & $\w3m$ & $\pbp$ \\
    \hline\hline
    5.25 & O & 200-800  & 0.82 & 0.97(2)  & 0.4289(5) & 0.027(2) & 0.01000(2) \\
    5.35 & O & 400-800  & 0.68 & 0.98(4)  & 0.451(3)  & 0.035(4) & 0.01000(9) \\
    5.45 & D & 400-800  & 0.74 & 1.09(5)  & 0.4769(8) & 0.049(7) & 0.00985(7) \\
    5.55 & D & 600-1200 & 0.80 & 1.00(3)  & 0.531(1)  & 0.175(7) & 0.00718(7) \\
    5.65 & D & 400-800  & 0.79 & 0.96(2)  & 0.5507(9) & 0.216(4) & 0.00677(2) \\
    5.75 & D & 200-800  & 0.88 & 0.989(7) & 0.5663(4) & 0.243(3) & 0.00658(1)
  \end{tabular}
  \caption{Data for an $8^3\times 4$ lattice with $m_0=1.65$, $L_s=12$, and $m_f=0.1$.}
  \label{tab:8nt4_h1.65_beta_crit}
\end{table}

\begin{table}
  \centering
  HMC traj.~len: $\frac{1}{50}\times 25$, \qquad CG stop cond: $10^{-6}$
  \begin{tabular}{cccccccc}
    $\beta$ & start & \# traj & acc. &
    $\left\langle e^{-\Delta H}\right\rangle$ & $\plaq$ & $\w3m$ & $\pbp$ \\
    \hline\hline
    5.15 & O & 200-800 & 0.83 & 1.02(2) & 0.4191(8) & 0.029(1) & 0.01485(5) \\
    5.25 & O & 400-800 & 0.66 & 0.97(5) & 0.4381(6) & 0.038(2) & 0.01458(5) \\
    5.35 & O & 400-800 & 0.63 & 0.97(5) & 0.471(2)  & 0.052(3) & 0.0134(2)  \\
    5.45 & O & 400-800 & 0.76 & 1.01(3) & 0.515(2)  & 0.161(4) & 0.0097(1)  \\
    5.55 & D & 400-800 & 0.79 & 1.05(5) & 0.540(1)  & 0.200(9) & 0.0088(1)  \\
    5.65 & D & 200-800 & 0.89 & 1.01(2) & 0.5570(5) & 0.242(4) & 0.00828(2)
  \end{tabular}
  \caption{Data for an $8^3\times 4$ lattice with $m_0=1.8$, $L_s=12$, and $m_f=0.1$.}
  \label{tab:8nt4_h1.8_beta_crit}
\end{table}

\begin{table}
  \centering
  CG stop cond: $10^{-6}$
  \begin{tabular}{ccccccccc}
    $\beta$ & start & traj.~len. & \# traj & acc. &
    $\left\langle e^{-\Delta H}\right\rangle$ & $\plaq$ & $\w3m$ & $\pbp$ \\
    \hline\hline
    5.0  & O & $\frac{1}{40}\times 20$ & 200-800  & 0.37 & 0.8(1)  & 0.4002(8) &
      0.032(2) & 0.01919(5) \\
    5.2  & O & $\frac{1}{40}\times 20$ & 400-800  & 0.43 & 1.2(3)  & 0.437(1)  &
      0.032(2) & 0.01838(4) \\
    5.25 & O & $\frac{1}{50}\times 25$ & 400-800  & 0.65 & 1.10(9) & 0.452(1)  &
      0.049(6) & 0.0174(2)  \\
    5.35 & D & $\frac{1}{50}\times 25$ & 600-1200 & 0.69 & 0.95(5) & 0.493(2)  &
      0.107(9) & 0.0135(4)  \\
    5.45 & D & $\frac{1}{50}\times 25$ & 600-1200 & 0.74 & 1.01(4) & 0.528(1)  &
      0.197(4) & 0.01039(7) \\
    5.55 & D & $\frac{1}{50}\times 25$ & 400-830  & 0.82 & 1.00(1) & 0.5463(5) &
      0.227(6) & 0.00974(4) \\
    5.65 & D & $\frac{1}{50}\times 25$ & 400-800  & 0.88 & 1.03(1) & 0.5613(8) &
      0.248(5) & 0.00943(4)
  \end{tabular}
  \caption{Data for an $8^3\times 4$ lattice with $m_0=1.9$, $L_s=12$, and $m_f=0.1$.}
  \label{tab:8nt4_h1.9_beta_crit}
\end{table}

\begin{table}
  \centering
  HMC traj.~len: $\frac{1}{50}\times 25$, \qquad CG stop cond: $10^{-6}$
  \begin{tabular}{cccccccc}
    $\beta$ & start & \# traj & acc. &
    $\left\langle e^{-\Delta H}\right\rangle$ & $\plaq$ & $\w3m$ & $\pbp$ \\
    \hline\hline
    5.05 & O & 200-800  & 0.77 & 0.99(3) & 0.4192(8) & 0.035(1) & 0.02324(8) \\
    5.15 & O & 200-800  & 0.75 & 0.98(3) & 0.442(1)  & 0.042(3) & 0.0215(2)  \\
    5.25 & O & 200-1200 & 0.79 & 1.03(1) & 0.474(1)  & 0.080(7) & 0.0181(3)  \\
    5.35 & D & 200-800  & 0.83 & 1.00(2) & 0.5130(7) & 0.173(6) & 0.0130(2)  \\
    5.45 & D & 200-800  & 0.87 & 1.02(2) & 0.5349(5) & 0.203(3) & 0.01157(4) \\
    5.55 & D & 200-800  & 0.85 & 1.01(1) & 0.5503(4) & 0.235(3) & 0.01099(2)
  \end{tabular}
  \caption{Data for an $8^3\times 4$ lattice with $m_0=2.0$, $L_s=12$, and $m_f=0.1$.}
  \label{tab:8nt4_h2.0_beta_crit}
\end{table}

\begin{table}
  \centering
  HMC traj.~len: $\frac{1}{50}\times 25$, \qquad CG stop cond: $10^{-6}$
  \begin{tabular}{cccccccc}
    $\beta$ & start & \# traj & acc. &
    $\left\langle e^{-\Delta H}\right\rangle$ & $\plaq$ & $\w3m$ & $\pbp$ \\
    \hline\hline
    4.85 & O & 200-800  & 0.69 & 0.95(3) & 0.4004(8) & 0.034(2) & 0.0323(2)  \\
    4.95 & O & 200-800  & 0.72 & 0.97(3) & 0.419(2)  & 0.040(2) & 0.0302(3)  \\
    5.05 & O & 200-800  & 0.48 & 0.92(4) & 0.443(2)  & 0.052(3) & 0.0272(5)  \\
    5.15 & O & 200-1200 & 0.62 & 1.01(3) & 0.480(3)  & 0.12(1)  & 0.0203(7)  \\
    5.25 & O & 400-800  & 0.70 & 0.97(5) & 0.5105(4) & 0.185(2) & 0.01559(8) \\
    5.35 & D & 400-800  & 0.69 & 1.01(4) & 0.529(1)  & 0.216(6) & 0.0141(1)  \\
    5.45 & D & 400-800  & 0.71 & 1.00(3) & 0.5453(7) & 0.230(4) & 0.01330(5)
  \end{tabular}
  \caption{Data for an $8^3\times 4$ lattice with $m_0=2.15$, $L_s=12$, and $m_f=0.1$.}
  \label{tab:8nt4_h2.15_beta_crit}
\end{table}

\begin{table}
  \centering
  HMC traj.~len: $\frac{1}{50}\times 25$, \qquad CG stop cond: $10^{-6}$
  \begin{tabular}{cccccccc}
    $\beta$ & start & \# traj & acc. &
    $\left\langle e^{-\Delta H}\right\rangle$ & $\plaq$ & $\w3m$ & $\pbp$ \\
    \hline\hline
    4.65 & O & 100-800 & 0.63 & 1.02(5) & 0.3953(6) & 0.046(3) & 0.0484(3)  \\
    4.75 & O & 200-800 & 0.68 & 0.99(5) & 0.4156(9) & 0.054(3) & 0.0442(3)  \\
    4.85 & O & 300-800 & 0.70 & 0.94(4) & 0.439(2)  & 0.069(5) & 0.0380(6)  \\
    4.95 & O & 200-800 & 0.77 & 1.02(4) & 0.4779(6) & 0.155(4) & 0.0257(2)  \\
    5.05 & D & 200-800 & 0.80 & 1.01(2) & 0.4987(9) & 0.190(2) & 0.0220(2)  \\
    5.15 & D & 200-800 & 0.84 & 1.01(3) & 0.5170(5) & 0.221(3) & 0.01962(7)
  \end{tabular}
  \caption{Data for an $8^3\times 4$ lattice with $m_0=2.4$, $L_s=12$, and $m_f=0.1$.}
  \label{tab:8nt4_h2.4_beta_crit}
\end{table}

\begin{table}
\begin{center}
  \begin{tabular}{llll}
    $L_s$ & $m_{\rm res}$ & $-b_0$ & $\chi^2/{\rm dof}$ \\
    \hline \hline
    10  	&     0.149(5)& 0.0094(5)     & 0.8(4) \\
    12      &     0.129(2)& 0.0080(2)     & 1.6(4) \\
    16      &     0.113(3)& 0.0080(4)     & 1.1(5) \\
    24	&	0.095(2)& 0.0075(3)	& 1.5(7) \\
    32	&	0.078(2)& 0.0068(5)	& 0.7(4) \\
    40      &     0.059(3)& 0.0048(3)     & 1.7(9) \\
  \end{tabular}
  \caption{ Values for $m^{(\rm GMOR)}_{\rm res}$ and $-b_0$ versus
  $L_s$ from fits to valence quark data with the dynamical quark
  mass fixed at $m_f = 0.02$.}
  \label{tab:mres_vs_ls}
\end{center}
\end{table}

\end{document}